\documentclass[twocolumn]{aa}
\usepackage{graphicx, amsmath, color, mathrsfs}
\graphicspath{{figures/},{figures/},{./}}
\usepackage{txfonts}
\usepackage{natbib,twoopt}
\usepackage[breaklinks=true]{hyperref} %% to avoid \citeads line fills
\bibpunct{(}{)}{;}{a}{}{,}             %% natbib format for A&A and ApJ
\makeatletter

\newcommandtwoopt{\citeads}[3][][]{\href{http://adsabs.harvard.edu/abs/#3}
	{\def\hyper@linkstart##1##2{}
	\let\hyper@linkend\@empty\citealp[#1][#2]{#3}}}

\newcommandtwoopt{\citepads}[3][][]{\href{http://adsabs.harvard.edu/abs/#3}
	{\def\hyper@linkstart##1##2{}
	\let\hyper@linkend\@empty\citep[#1][#2]{#3}}}

\newcommandtwoopt{\citetads}[3][][]{\href{http://adsabs.harvard.edu/abs/#3}
	{\def\hyper@linkstart##1##2{}
	\let\hyper@linkend\@empty\citet[#1][#2]{#3}}}

\makeatother
\usepackage{hyperref}
\usepackage{hypcap}
\definecolor{cyan}{cmyk}{1.,0.,0.,0.2}
\definecolor{vert}{cmyk}{0.5,0.,0.5,0.5}
\definecolor{magenta}{cmyk}{0.,1.,0.,0.1}
\definecolor{verdatre}{cmyk}{0.5,0.,0.5,0.5}
\definecolor{vert_clair}{cmyk}{0.5,0.,0.5,0.2}
\definecolor{yellow}{cmyk}{0.,0.,1.,0.0}
\definecolor{yellow_1}{cmyk}{0.,0.,0.5,0.0}
\definecolor{rouge}{cmyk}{0.,0.4,0.6,0.0}
\definecolor{orange}{cmyk}{0.,0.5,0.5,0.05}
\definecolor{violet}{rgb}{0.5,0.,0.5}
\definecolor{darwin_box}{rgb}{0.988,0.878,0.77}
\definecolor{darwin_text}{rgb}{0.1,0.07,0.02}
\newcommand{\beq}{\begin{equation}}
\newcommand{\eeq}{\end{equation}}
\newcommand{\bea}{\begin{eqnarray}}
\newcommand{\ena}{\end{eqnarray}}

\begin{document}
%%%%%%%%%%%%%%%%%%%%%%%%%%%%%%%%%%%%%%%%%%%%%%%%%%%%%%%%%%
%%%%%%%%%%%%%%%%%%%%%%%%%%%%%%%%%%%%%%%%%%%%%%%%%%%%%%%%%%
\title{Theoretical uncertainties in extracting cosmic-ray diffusion parameters: the boron-to-carbon ratio}

\author{Y.~Genolini\thanks{Contact author: yoann.genolini@lapth.cnrs.fr} \and A.~Putze \and P.~Salati \and P.~D.~Serpico
}

\institute{
LAPTh, Universit\'e Savoie Mont Blanc \& CNRS, 9 Chemin de Bellevue, B.P.110 Annecy-le-Vieux, F-74941, France
}
%

%\date{Received September 16, 2014; accepted}
\date{Received; accepted\\
Preprint numbers : LAPTH-018/15}
%

%%%%%%%%%%%%%%%%%%%%%%%%%%%%%%%%%%%%%%%%%%%%%%%%%%%%%%%%%%
%%%%%%%%%%%%%%%%%%%%%%%%%%%%%%%%%%%%%%%%%%%%%%%%%%%%%%%%%%
%
% \abstract{}{}{}{}{} -- 5 {} token are mandatory
%
\abstract
{PAMELA and, more recently, AMS-02, are ushering us into a new era of greatly reduced statistical uncertainties in experimental measurements of cosmic-ray fluxes. In particular, new determinations  of
traditional diagnostic tools such as the
boron-to-carbon ratio (B/C) are expected to significantly reduce errors on cosmic-ray diffusion parameters, with important
implications for astroparticle physics, ranging from inferring primary source spectra to indirect dark matter searches.
}
{
It is timely to stress, however, that the conclusions obtained crucially depend on the framework in which the data are interpreted as well as on some nuclear input parameters. We aim at assessing the theoretical uncertainties affecting the outcome, with models as simple as possible while still retaining the key dependencies.}
{
We compared different semi-analytical, two-zone model descriptions of cosmic-ray transport in the Galaxy: infinite slab(1D), cylindrical symmetry(2D) with homogeneous sources, cylindrical symmetry(2D) with inhomogeneous source distribution. We tested for the effect
of a primary source contamination in the boron flux by parametrically altering its flux. We also tested for nuclear cross-section uncertainties. All hypotheses were compared by $\chi^{2}$ minimisation techniques to preliminary results from AMS-02.
}
{
We find that the main theoretical bias on the determination of the diffusion coefficient index $\delta$ (up to a factor two) is represented by the assumption that no injection of boron takes place at the source. The next most important uncertainty is represented by cross-section uncertainties, which reach $\pm 20\%$ in $\delta$. As a comparison, nuclear uncertainties
are more important than the shift in the best-fit when introducing a convective wind of velocity $\lesssim$30 km/s, with
respect to a pure diffusive baseline model. Perhaps surprisingly, homogeneous 1D vs. 2D performances are similar in determining diffusion parameters. An inhomogeneous source distribution marginally alters the central value of the diffusion coefficient normalisation (at the 10\%, $1\,\sigma$ level). However, the index of the diffusion coefficient $\delta$ is basically unaltered, as well as the goodness of fit.
}
{
Our study suggests that, differently for instance from the leptonic case, realistic modelling of the geometry of the Galaxy
and  of the source distribution are of minor importance to correctly reproduce B/C data at high energies and thus,
to a large extent, for the extraction of diffusion parameters.
The ansatz on the lack of primary injection of boron represents the most serious bias and requires multi-messenger
studies to be addressed. If this uncertainty could be lifted, nuclear uncertainties would still represent a serious concern; they degrade the systematic error on the inferred parameters to the 20\% level, or three times the estimated experimental sensitivity. To reduce this, a new nuclear cross-section measurement campaign might be required.
}

\keywords{Astroparticle physics - cosmic rays - boron-to-carbon ratio - diffusion parameters}
\maketitle
%
%%%%%%%%%%%%%%%%%%%%%%%%%%%%%%%%%%%%%%%%%%%%%%%%%%%%%%%%%%
%%%%%%%%%%%%%%%%%%%%%%%%%%%%%%%%%%%%%%%%%%%%%%%%%%%%%%%%%%
%
\section{Introduction}

The pattern of relative abundances of nuclei in the cosmic radiation (CR) is roughly similar to the one of the solar system material, with some notable exceptions: fragile nuclei (with low binding energies) such as ${}^2$H or Li-Be-B are over-represented in CR. This CR component is usually interpreted as the result of production by spallation of heavier species during the propagation of primary cosmic rays---whose injected abundance is assumed to closely trace that of the solar system ---in the interstellar medium. The ratios of these secondary to primary fluxes have long been recognised as a tool for constraining CR propagation parameters, for some review see
for instance~\citet{Maurin:2002ua} and~\citet{Strong:2007nh}. The boron-to-carbon ratio, or B/C, represents the most notable example among them. The key constraints on the diffusion parameters are inferred by its measurement, with the corresponding confidence levels (see for instance~\citet{Maurin:2001sj}) widely used as benchmarks. It has also been recognised that datasets available one decade ago were still insufficient for a satisfactory measurement of these parameters~\citep{Maurin:2001sj}.

The current decade is undergoing a major shift, however, with experiments such as PAMELA\citepads{2014PhR...544..323A} and most notably AMS-02 (\url{http://www.ams02.org}), which are characterised by significantly increased precision and better control of systematics. The current and forthcoming
availability of high-quality data prompts the question of how best to exploit them to extract meaningful (astro)physical
information.
This new situation demands reassessing theoretical uncertainties, which will probably be the limiting factor in the parameter extraction accuracy. As a preliminary work, preceding the actual data analysis, we revisit this issue to determine the relative importance of various effects: some have already been considered in the past, some were apparently never quantified. We also found that the main theoretical biases or errors are related to phenomena that can be described in a very simple 1D diffusive model. We thus adopt it as a benchmark for our description, reporting the key formulae that thus have a pedagogical usefulness, too.
In fact, we focus on determining the diffusion coefficient, which we parameterise as conventionally in the literature (see for example \citetads{1997A&A...321..434P}):
\begin{equation}
D \left({\cal R} \right) = D_0 \, \beta \left(\frac{{\cal R}}{{\cal R}_0=1\,\text{GV}} \right)^\delta ,
\label{DofE}
\end{equation}
where $D_0$ and $\delta$ are determined by the level and power-spectrum of hydromagnetic turbulences, ${\cal R}$ is the rigidity, and the velocity $\beta=v/c\simeq 1$ in the high-energy regime of interest here (kinetic energy/nucleon $\gtrsim 10\,$GeV/nuc).
In fact, at lower energies numerous effects, in principle of similar magnitude, are present, such as convective winds, reacceleration, and collisional losses. At high energy, there is a common consensus that only diffusion and source-related effects
are important. We focus on the high-energy region since it is the cleanest to extract diffusion parameters, that is the least subject to parameter degeneracies. While adding lower-energy data can lead to better constraints from a statistical point of view, the model dependence cannot but grow.
Since our purpose is to compare theoretical with statistical uncertainties from observations, our choice is thus conservative: in a global analysis, the weight of the former with respect to the latter is probably larger.
To deal with a realistic level of statistical errors of the data that will be available for the forthcoming analyses, we base our analyses on preliminary AMS-02 data of the B/C ratio~\citep{2013..ICRC}.

This paper is organised as follows. In Sect.~\ref{1dmodel1} we recall a simple 1D diffusion model providing our benchmark for the following analyses. This model certainly has pedagogical value, since it allows encoding the main dependences of the B/C ratio on input as well as astrophysical parameters in simple analytical formulae
. At the same time,
it provides a realistic description of the data, at least if one limits the analysis to sufficiently high energies.
Relevant formulae are introduced in Sect.~\ref{formulae1}, while in subsection~\ref{fitting} we recall the main statistical
tools used for the analysis. In Sect.~\ref{PrimaryBoron} we describe the main degeneracy affecting the analysis: the one with possible injection of boron nuclei at the sources. The next most important source of error is associated to
cross-section uncertainties, to which we devote Sect.~\ref{sigmas}. In Sect.~\ref{propmodeling} we discuss
relatively minor effects linked to modelling of the geometry of the diffusion volume, source distribution,
or the presence of convective winds. In Sect.~\ref{conclusions} we report our conclusions.

%%%%%%%%%%%%%%%%%%%%%%%%%%%%%%%%%%%%%%%%%%%%%%%%%%%%%%%%%%
%%%%%%%%%%%%%%%%%%%%%%%%%%%%%%%%%%%%%%%%%%%%%%%%%%%%%%%%%%
\section{B/C fit with a 1D model}
\label{1dmodel1}
%%%%%%%%%%%%%%%%%%%%%%%%%%%%%%%%%%%%%%%%%%%%%%%%%%%%%%%%%%
%%%%%%%%%%%%%%%%%%%%%%%%%%%%%%%%%%%%%%%%%%%%%%%%%%%%%%%%%%
%
%%%%%%%%%%%%%%%%%%%%%%%%%%%%%%%%%%%%%%%%%%%%%%%%%%%%%%%%%%
\subsection{1D diffusion model}\label{formulae1}
%FFFFFFFFFFFFFFFFFFFFFFFFFFFFFFFFFFFFFFFFFFFFFFFFFFFFFFFFFFFFFFFFFFFFFFFFFFFFFFFFFFFFF
\begin{figure}[!ht]
\centering
\includegraphics[width=0.45\textwidth]{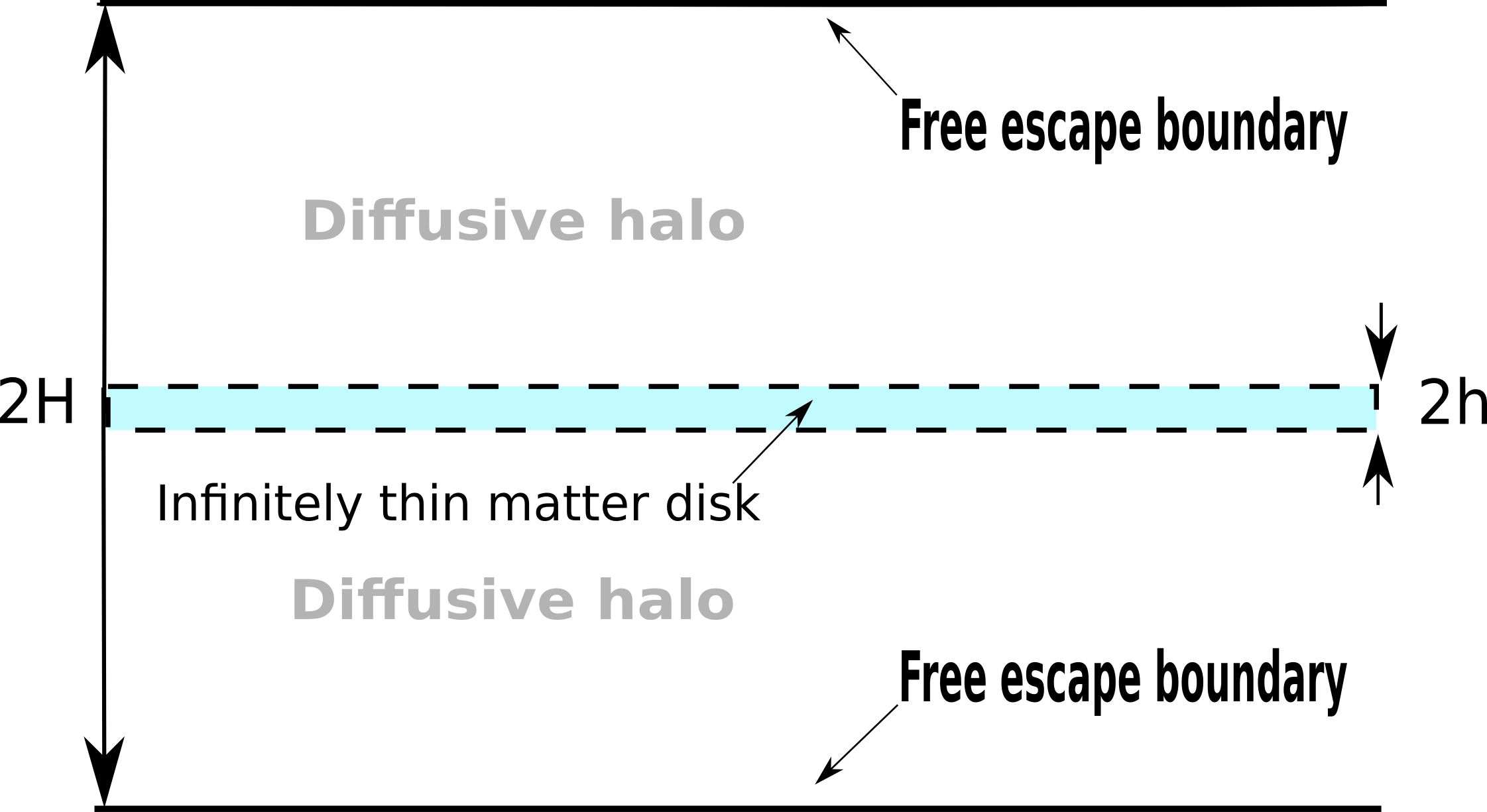}
\caption{Sketch of the 1D slab model of the Galaxy, with matter homogeneously distributed inside an infinite plane of thickness $2h$
sandwiched between two thick diffusive layers  of thickness $2H$.}
\label{fig:1}
\end{figure}
%FFFFFFFFFFFFFFFFFFFFFFFFFFFFFFFFFFFFFFFFFFFFFFFFFFFFFFFFFFFFFFFFFFFFFFFFFFFFFFFFFFFFF

The simplest approach to model the transport of cosmic-ray nuclei inside the Galaxy is to assume that their production is confined
inside an infinite plane of thickness $2h$, that is sandwiched inside an infinite diffusion volume of thickness $2H$, symmetric above and below the plane.
The former region
stands for the Galactic disk, which comprises the gas and the massive stars of the Milky Way, whereas the latter domain represents
its magnetic halo. A sketch of this model is given in Fig.~\ref{fig:1}.
The boundary conditions fix the density of cosmic rays at the halo edges $z = \pm H$ to zero, while the condition $h \ll H$ (in practice,
$h$ is almost two orders of magnitude smaller than $H$) allows us to model the Galactic matter distribution as an infinitely thin disk
whose vertical distribution is accounted for by the Dirac function $2 h \delta(z)$.
Our focus on energies above 10\,GeV/nuc allows us to neglect continuous (ionisation and Coulomb) energy losses, electronic
capture, and reacceleration. These subleading effects cannot be truly considered as theoretical uncertainties, since they can be
introduced by a suitable upgrade of the model. However, taking them into account at this stage would imply a significant loss in simplicity
and transparency.

The well-known propagation equation for the (isotropic part of the gyrophase-averaged) phase space density $\psi_a$ of a stable nucleus $a$, with
charge (atomic number) $Z_a$,  expressed in units of {particles cm$^{-3}$ (GeV/nuc)$^{-1}$}, takes the form
\begin{align}
\frac{\partial \psi_a}{\partial t}  -
\frac{\partial}{\partial z} \left( D \frac{\partial \psi_a}{\partial z} \right) =
& \,2 h \delta(z) \cdot q_{a} +
\delta(z) \sum_{Z_b \geqslant Z_a}^{Z_{max}}
\sigma_{b \to a} \cdot  v \frac{\mu}{m_{\rm ISM}} \psi_b  \nonumber \\&-
\delta(z) \cdot  \sigma_{a} \cdot v \frac{\mu}{m_{\rm ISM}} \psi_a ,
\label{eqprop2}
\end{align}
where the spatial diffusion coefficient $D$ has been defined in Eq.~(\ref{DofE}).
The cross-section for the production of the species $a$ from the species $b$ through its interactions with the interstellar medium (ISM)
is denoted by $\sigma_{b \to a}$, whereas $\sigma_{a}$ is the total inelastic interaction (destruction) cross-section of the species $a$ with the ISM.
The fragmentation of the nucleus $b$ takes place at constant energy per nucleon. $v$ hence stands for the velocities of both parent
($b$) and child ($a$) nuclei.
The surface density of the Galactic disk is denoted by $\mu$, while $m_\text{ISM}$ is the average mass of the atomic gas that it contains.
The values of the production cross-sections $\sigma_{b \to a}$ were calculated with the most recent formulae from \citetads{2003ApJS..144..153W}.
The destruction cross-sections $\sigma_{a}$ were computed by the semi-empirical formulae of \citetads{1997lrc..reptQ....T, 1999NIMPB.155..349T}.
The high-energy shapes of both cross-sections exhibit a plateau that allows one to approximate them as constants in this energy range.

\vskip 0.1cm
Solving the propagation Eq.~(\ref{eqprop2}) in the steady-state regime allows expressing the flux ${\cal J}_a \equiv ({v}/{4 \pi}) \, \psi_a$
of the stable nucleus~$a$ inside the Galactic disk ($z=0$) as
\begin{align}
{\cal J}_a (E_k)  &=  \left\{
{Q_a + \sum_{Z_b \geqslant Z_a}^{Z_{\text{max}}}} \sigma_{b \to a} \cdot  {\cal J}_b \right\} /
\left\{ {\sigma^{\text{diff}} + \sigma_{a}} \right\}, \label{eq:flux} \\ \nonumber
\text{where} \quad
\sigma^{\text{diff}} &= \frac{2 D \, m_{\text{ISM}}}{\mu v H}.
\end{align}
The fluxes ${\cal J}_b$ of the parent species are also taken at $z=0$. Here $Q_a$, which stands for the source term, is homogeneous to a flux times
a surface and is expressed in units of {particles (GeV/nuc)$^{-1}$ s$^{-1}$ sr$^{-1}$}. It is related to $q_{a}$ through
\begin{equation}
Q_a  =
 \frac{1}{4 \pi} \cdot  \frac{q_a}{n_{\text{ISM}}} \equiv
N_a \, \left( \frac{\cal R}{1 \, \text{GV}} \right)^{\alpha} ,
\end{equation}
where $N_a$ is a normalisation constant that depends on the isotope $a$. We assumed an injection spectrum with the same spectral index $\alpha$
for all nuclei.
The value of $N_a$ should be adjusted by fitting the corresponding flux ${\cal J}_a$ to the measurements performed at Earth. However, these scarcely contain information on the isotopic composition of cosmic rays. Nuclei with the same charge $Z$ are in general collected together, irrespective of their mass. More isotopic observations would be necessary to set the values of the coefficients $N_a$ for the various isotopes $a$ of the same element.

\vskip 0.1cm
In our analysis, we assumed solar system values\citepads{2003ApJ...591.1220L} for the isotopic fractions $f_a$ of the stable species $a$ that were injected at the sources. We then proceeded by computing the flux ${\cal J}_Z$ of each element $Z$ at Earth. We fixed the normalisation $N_Z$ for the total injection of all stable isotopes of the same charge $Z$ by fitting the measured flux of that element.
The normalisation entering in the calculation of $Q_a$ is given by $N_a = f_a \cdot N_Z$, the sum of the fractions $f_a$ corresponding to the same element $Z$ amounting to 1.
The actual isotopic composition of the material accelerated at the source might be different from the solar system one, as is the case for neon\citepads{2008NewAR..52..427B}. Our method might introduce a theoretical bias in CR element flux calculations. However, our main focus here is to extract the propagation parameters thanks to the different sensitivities between primary carbon and (a priori) secondary boron.
The only isotopes that come into play in the B/C ratio are the stable nuclei $^{12}\text{C}$, $^{13}\text{C}$, $^{10}\text{B}$, and $^{11}\text{B}$, unstable $^{14}\text{C}$ plays a very minor role. The isotopes of either carbon or boron have similar rigidities and destruction cross-sections. Varying the isotopic composition of carbon (and of boron, should it be partially primary) does not affect the ratio calculation. Futhermore, secondary boron is mainly produced by the fragmentation of one particular isotope of each heavier element. For example, the primary component of $^{12}\text{C}$ is two orders of magnitude larger than that of $^{13}\text{C}$. This reduces the differences arising from the boron production cross-sections. Althought most of the isotopes at stake are stable, radioactive nuclei were also taken into account in the calculation, and we obtained more complicated expressions for the fluxes, which are not displayed here for brevity.
They are reported for instance in Appendix A of~\cite{Putze:2010zn}.
By defining the total flux of a nucleus of charge $Z$ as the sum over all its isotopes $a$
\begin{equation}
{\cal J}_Z = \sum_{\text{isotopes $a$} \atop \text{of same }Z} {\cal J}_{a} ,
\end{equation}
and considering only the dominant contribution from stable nuclei, the B/C flux ratio can be written as
\begin{equation}
\frac{{\cal J}_\text{B} (E_k)}{{\cal J}_\text{C} (E_k)}  =  \left\{
\frac{Q_\text{B}}{{\cal J}_\text{C}} + \sigma_{\text{C} \to \text{B}} +
\sum_{ Z_b  > Z_{\text{C}}}^{ Z_\text{max}} \sigma_{b \to \text{B}} \cdot
\frac{{\cal J}_{b}}{{\cal J}_\text{C}} \right\} /
\left\{ {\sigma^{\text{diff}} + \sigma_\text{B}} \right\} .
\label{eq:b_to_c_primary_B}
\end{equation}
If we assume that there are no primary boron sources, that is, $Q_\text{B} = 0$, this expression simplifies into
\begin{equation}
\frac{{\cal J}_\text{B} (E_k)}{{\cal J}_\text{C} (E_k)} =
\frac{\sigma_{\text{C} \to \text{B}}}{\sigma^{\text{diff}} + \sigma_{B}} +
\sum_{Z_b  >  Z_\text{C}}^{Z_\text{max}} \frac{\sigma_{b \to \text{B}}}{\sigma^{\text{diff}} + \sigma_\text{B}}
\cdot \frac{{\cal J}_{b}}{{\cal J}_\text{C}}.
\label{eq:b_to_c}
\end{equation}
The impact of relaxing this hypothesis is explored in Sect.~\ref{PrimaryBoron} where the effect of a non-vanishing value for $Q_\text{B}$
is considered.

%
%%%%%%%%%%%%%%%%%%%%%%%%%%%%%%%%%%%%%%%%%%%%%%%%%%%%%%%%%%
\subsection{Fitting procedure and benchmark values for this study}\label{fitting}

We used the AMS-02 recent release of the B/C ratio~\citep{2013..ICRC}
to study the impact of systematics on the propagation parameters. As explained above, we limited ourselves to the high-energy sub-sample, above 10 GeV/nuc.
The set of Eqs.~(\ref{eq:flux}) is of triangular form. The heaviest element considered in the network, which in our case is $^{56}$Fe, can only suffer
destruction. No other heavier species $b$ enters in the determination of its flux ${\cal J}_a$, which hence is proportional to the injection term $Q_a$.
Once solved for it, the algebraic relation yields the solution for the lighter nuclei, down to boron. We evaluated the cascade down to beryllium to take into account its radioactive decay into boron.

\vskip 0.1cm
The primary purpose of our analysis is to determine the diffusion parameters $D_0$ and $\delta$ from the B/C flux ratio
${\cal F} \equiv {{\cal J}_\text{B}}/{{\cal J}_\text{C}}$.
Another parameter of the model is the magnetic halo thickness $H$. As shown in Eq.~(\ref{eq:flux}), $D_{0}$ and $H$ are completely degenerate
when only considering stable nuclei, which provide the bulk of cosmic rays. In the following, $H$ is therefore fixed at $4$\,kpc for simplicity,
although it should be kept in mind that, to a large extent, variations in $D_0$ can be traded for variations in $H$.
Finally, the injection spectral index $\alpha$ also enters in the calculation of the B/C ratio through the source terms $Q_a$. How strong its effect is on the best-fit diffusion parameters $D_0$ and $\delta$ is one of the questions we treat in this section.
To this purpose, we carried out a chi-square ($\chi^2$) analysis of the B/C observations and minimised the function
\begin{equation}
\chi^{2}_{\text{B/C}} = \sum_i \left\{
\frac{{\cal F}^{\text{exp}}_i - {\cal F}^{\text{th}}_i \left(\alpha , \delta , D_0 \right)}{\sigma_{i}} \right\}^{2} ,
\end{equation}
where the sum runs over the data points $i$ whose kinetic energies per nucleon are $E_{k,i}$, while ${\cal F}^{\rm exp}_{i}$ and $\sigma_{i}$
stand for the central values and errors of the measurements. The theoretical expectations ${\cal F}^{\rm th}_{i}$ also depend on the normalisation
constants $N_a$, which come into play in the source terms $Q_a$ of the cascade relations~(\ref{eq:flux}). To determine them, we first fixed the spectral
index $\alpha$ and the diffusion parameters $D_0$ and $\delta$. We then carried out an independent $\chi^2$-based fit on the fluxes ${\cal J}_Z$ of
the various elements that belong to the chain that reaches from iron to beryllium. The measured fluxes are borrowed from the cosmic-ray
database of\citetads{2014A&A...569A..32M} from which we selected the points above 10 GeV/nuc. As explained above, this method yields the constants
$N_Z$ and eventually the values of $N_a$ once the solar system isotopic fractions $f_a$ are taken into account. The overall procedure amounts
to profile over the normalisation constants $N_a$ to derive $\chi^{2}_{\rm B/C}$ as a function of $\alpha$, $\delta$ and $D_0$.
Minimisations were performed by MINUIT (\url{http://www.cern.ch/minuit}), a package interfaced in the ROOT programme (\url{https://root.cern.ch}).

%FFFFFFFFFFFFFFFFFFFFFFFFFFFFFFFFFFFFFFFFFFFFFFFFFFFFFFFFFFFFFFFFFFFFFFFFFFFFFFFFFFFFF
\begin{figure}[!ht]
\centering
\includegraphics[width=0.5\textwidth]{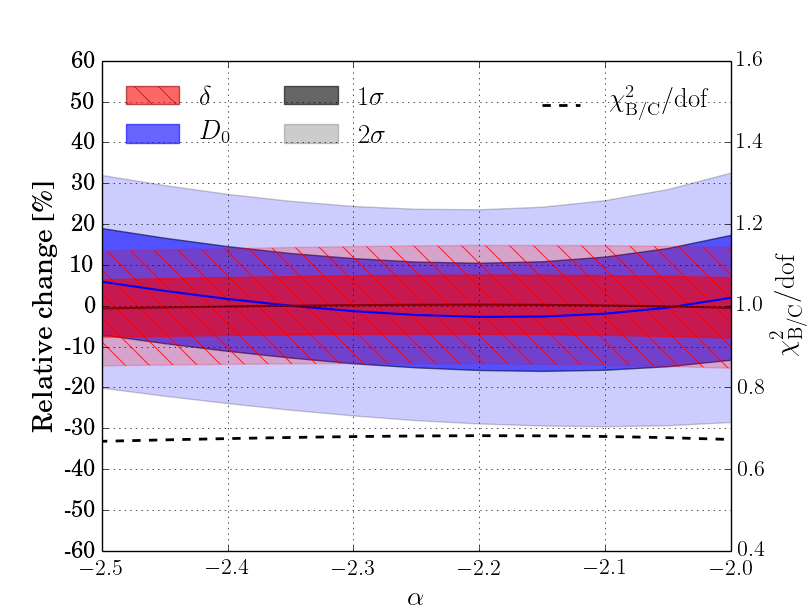}
\caption{Relative variations of the best-fit propagation parameters (compared to the benchmark model of Table~\ref{tab:benchmark})
with respect to the injection spectral index $\alpha$.}
\label{fig:2}
\end{figure}
%FFFFFFFFFFFFFFFFFFFFFFFFFFFFFFFFFFFFFFFFFFFFFFFFFFFFFFFFFFFFFFFFFFFFFFFFFFFFFFFFFFFFF

\vskip 0.1cm
To check the accuracy and robustness of our fitting procedure, a preliminary test is in order.
A commonly accepted notion is that the B/C ratio does not depend, to leading order, on the spectral index $\alpha$.
There is indeed no dependence on $\alpha$ in the cross-section ratios of Eq.~(\ref{eq:b_to_c}) in the pure diffusive regime where
$\sigma_\text{B} \ll \sigma^{\rm diff}$. We have checked numerically that this behaviour holds by calculating the B/C best-fit
values of the diffusion parameters at fixed spectral index $\alpha$. The results are reported in Fig.~\ref{fig:2}, where $D_0$ and $\delta$
are plotted, with their confidence limits, as a function of $\alpha$. We scanned over the physical range that extends from $-2.5$ to
$-2$ and observed that the relative variations of $D_0$ and $\delta$ are 5\% and 1\%, respectively.
The blue ($D_0$) and red ($\delta$) bands are almost horizontal. An anti-correlation between $D_0$ and $\delta$ is marginally noticeable and
can be understood by the interplay of these parameters inside the diffusion coefficient $D$, the only relevant parameter that the B/C
fit probes. We attribute the small variation of $D_0$ with $\alpha$ to the different sensitivities of the normalisation
constants $N_Z$ of nitrogen and oxygen to the low-energy data points as compared to carbon. This could result in fluctuations
of the ${N_\text{N}}/{N_\text{C}}$ and ${N_\text{O}}/{N_\text{C}}$ ratios with respect to the actual values.
In any case, the extremely small dependence of the B/C ratio on $\alpha$ confirms the naive expectations and suggests that it is useless and simply impractical to keep $\alpha$ as a free parameter.

\begin{figure}[!h]
\centering
\includegraphics[width=0.5\textwidth]{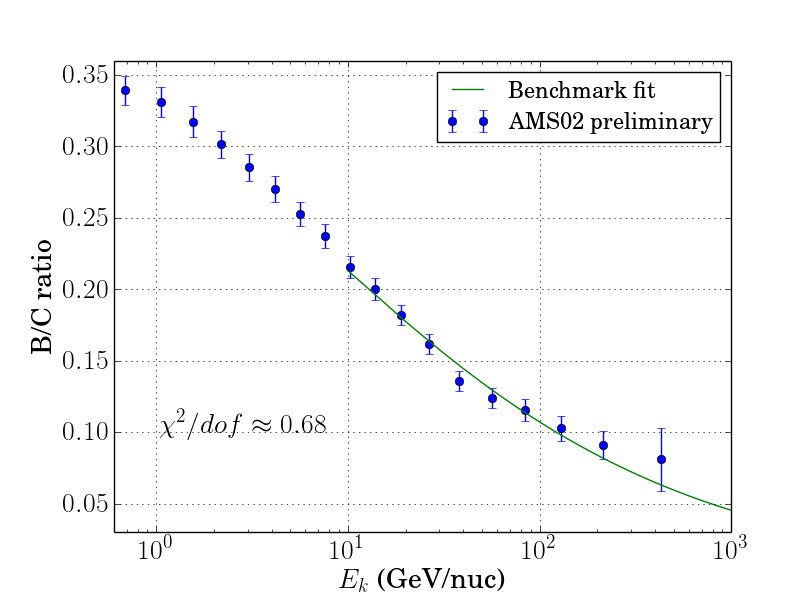}
\caption{
Preliminary AMS-02 measurements of the B/C ratio~\citep{2013..ICRC} are plotted as a function of kinetic energy
per nucleon. The theoretical prediction of the 1D/slab reference model of Tab.~\ref{tab:benchmark} is also featured for
comparison.}
\label{fig:fig:3}
\end{figure}

\vskip 0.1cm
Nonetheless, there is a particular value of the injection index that best fits the fluxes of the elements $Z$ that come into play in the cascade from iron to beryllium. By minimising the $\chi^2$-function
\begin{equation}
\chi^2_{\cal J} =
\sum_{Z \geqslant Z_\text{Be}}^{Z_\text{Fe}}
\sum_i \left\{
\frac{{\cal J}^{\rm exp}_{Z,i}(E_{k,i} ) - {\cal J}^{\rm th}_{Z,i}(E_{k,i} )}{\sigma_{Z,i}}
\right\}^{2} ,
\end{equation}
we find $\alpha=-2.34$ as our benchmark value. Applying then our B/C analysis yields the propagation parameters $D_{0}$ and $\delta$ of the reference model
of Table~\ref{tab:benchmark} which we used for the following analyses.
The corresponding B/C ratio is plotted in Fig.~\ref{fig:fig:3} as a function of kinetic energy per nucleon and compared
to the preliminary AMS-02 measurements~\citep{2013..ICRC}.
In what follows, we study how $D_{0}$ and $\delta$ are affected by a few effects under scrutiny and gauge the magnitude of their changes with respect to the reference model. We could have decided to keep the injection index $\alpha$ equal to its fiducial value of -2.34, but we preferred to fix the spectral index $\gamma=\alpha-\delta=-2.78$ of the high-energy fluxes ${\cal J}_Z$ at Earth. Keeping $\alpha$ fixed would have little effect on the B/C ratio, but would degrade the goodness of the fits on absolute fluxes.

%TTTTTTTTTTTTTTTTTTTTTTTTTTTTTTTTTTTTTTTTTTTTTTTTTTTTTTTTTTTTTTTTTTTTTTTTTTTTT
\begin{table}[!h]
\begin{center}
\begin{tabular}{|l|c|}
\hline
\multicolumn{2}{|c|}{Reference parameter values} \\
\hline\hline
$\alpha$ & $-2.34$ \\
$D_{0}$ [kpc$^2$/Myr] & $(5.8 \pm 0.7) \cdot 10^{-2}$ \\
$\delta$ & $0.44 \pm 0.03$ \\
${\chi^{2}_{\rm B/C}}/{\rm dof}$ & $5.4/8 \approx 0.68$ \\
\hline
$\gamma=\alpha-\delta$ (fixed) & $-2.78$ \\
\hline
\end{tabular}
\vskip 0.2cm
\caption{Benchmark best-fit  parameters  of the 1D/slab model, with respect to which comparisons are subsequently made.}
\label{tab:benchmark}
\end{center}
\end{table}
%TTTTTTTTTTTTTTTTTTTTTTTTTTTTTTTTTTTTTTTTTTTTTTTTTTTTTTTTTTTTTTTTTTTTTTTTTTTTT

%FFFFFFFFFFFFFFFFFFFFFFFFFFFFFFFFFFFFFFFFFFFFFFFFFFFFFFFFFFFFFFFFFFFFFFFFFFFFFFFFFFFFF
%FFFFFFFFFFFFFFFFFFFFFFFFFFFFFFFFFFFFFFFFFFFFFFFFFFFFFFFFFFFFFFFFFFFFFFFFFFFFFFFFFFFFF

\begin{figure}[!ht]
\centering
\includegraphics[width=0.5\textwidth]{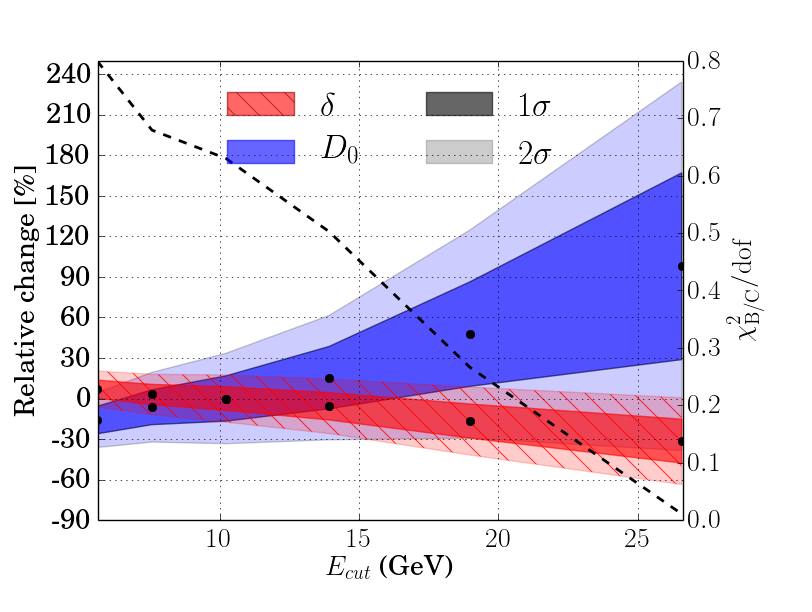}
\caption{Relative variations of the best-fit propagation parameters, as compared to the benchmark model of
Tab.~\ref{tab:benchmark}, with respect to the low-energy cut-off $E_{\rm cut}$ above which we carry out
the B/C analysis. }
\label{fig:4}
\end{figure}

\vskip 0.1cm
Another crucial test of our fitting procedure is to check how the results depend on the low-energy cut-off $E_{\rm cut}$ above
which we carried out our analysis. We set the flux spectral index $\gamma$ to its benchmark value of Table~\ref{tab:benchmark}
and determined the B/C best-fit values of the diffusion parameters as a function of $E_{\rm cut}$, which was varied from 5 to 30\,GeV/nuc. The results are plotted in Fig.~\ref{fig:4} with the $1{\sigma}$ and $2{\sigma}$ uncertainty bands.
As expected, the statistical errors increase when moving from a low $E_{\rm cut}$ to a higher value. That is why the reduced
$\chi^2$ (dashed line) decreases steadily as the cut-off energy is increased. The higher the cosmic-ray energy, the fainter the flux
and the scarcer the events in the detector. The widths of the blue ($D_0$) and red ($\delta$) bands at
$E_{\rm cut}=10$\,GeV/nuc, however, are not significantly larger than for a cut-off energy of 5~GeV/nuc. This suggests that our
estimates for the statistical errors are slightly pessimistic, which is acceptable and consistent with our purpose.

\vskip 0.1cm
The other trend that we observe in Fig.~\ref{fig:4} is a shift in the preferred value of $\delta$ to increasingly lower values as we limit
the analysis to increasingly higher energies. This is no limitation of our procedure. On the contrary, it is a real feature
that the data exhibit, as is clear in Fig.~\ref{fig:fig:3}, where the tail of the B/C points does look flatter above 50\,GeV/nuc. The
anti-correlation between $\delta$ and $D_0$ that we observe in Fig.~\ref{fig:4} has already been explained by the interplay of these
two parameters inside the diffusion coefficient $D$, to which the B/C ratio is sensitive. The increase of
$D_0$ is then generic and does not signal any new effect.
At that stage, the statistical uncertainties are still of the same order as the systematic uncertainties generated by using different energy cuts.
Should the decrease of $\delta$ with $E_{\rm cut}$ be confirmed with higher statistics, some intrinsic explanation might be necessary
for the failure of a power-law fit. See for instance Sect.~\ref{PrimaryBoron} for a possible explanation.

%FFFFFFFFFFFFFFFFFFFFFFFFFFFFFFFFFFFFFFFFFFFFFFFFFFFFFFFFFFFFFFFFFFFFFFFFFFFFFFFFFFFFF
%FFFFFFFFFFFFFFFFFFFFFFFFFFFFFFFFFFFFFFFFFFFFFFFFFFFFFFFFFFFFFFFFFFFFFFFFFFFFFFFFFFFFF

%%%%%%%%%%%%%%%%%%%%%%%%%%%%%%%%%%%%%%%%%%%%%%%%%%%%%%%%%%
%%%%%%%%%%%%%%%%%%%%%%%%%%%%%%%%%%%%%%%%%%%%%%%%%%%%%%%%%%
\section{Primary boron?}
\label{PrimaryBoron}
%%%%%%%%%%%%%%%%%%%%%%%%%%%%%%%%%%%%%%%%%%%%%%%%%%%%%%%%%%
%%%%%%%%%%%%%%%%%%%%%%%%%%%%%%%%%%%%%%%%%%%%%%%%%%%%%%%%%%

Typical fits of the B/C ratio are based on the assumption that no boron is accelerated at the source, so that the term proportional
to $Q_{\text B}$ at the right-hand side of Eq.~(\ref{eq:b_to_c_primary_B}) vanishes. However, this is just an assumption that need to be tested empirically. It is crucially linked to the hypothesis that the acceleration time is much shoter than the
propagation time within the magnetic halo and that it occurs in a low-density environment.
On the other hand, typical astrophysical accelerators such as supernova remnants might have the capability to accelerate up to TeV
energies for of about $t_{\rm life} \sim 10^{5}$ years in an interstellar medium with $n_{\rm ISM} \sim 1$~cm$^{-3}$, or greater
when surrounded by denser circumstellar material. The corresponding surface density
$n_{\rm ISM} \, c \, t_{\rm life} \sim 10^{23}$\,cm$^{-2}$
easily leads to percent-level probabilities for nuclei to undergo spallation in the sources. A factor of only a few times higher than this
would certainly have dramatic consequences on the information inferred from secondary-to-primary ratios. More elaborate versions of
this idea and related phenomenology have also been detailed as a possible explanation of the hard spectrum of secondary positron
data~\citep{Blasi:2009hv,Blasi:2009bd,Mertsch:2009ph}, which was recently compared with the AMS-02
data\citepads{2014PhRvD..90f1301M}.

%FFFFFFFFFFFFFFFFFFFFFFFFFFFFFFFFFFFFFFFFFFFFFFFFFFFFFFFFFFFFFFFFFFFFFFFFFFFFFFFFFFFFF
\begin{figure}[!htb]
\includegraphics[width=0.5\textwidth]{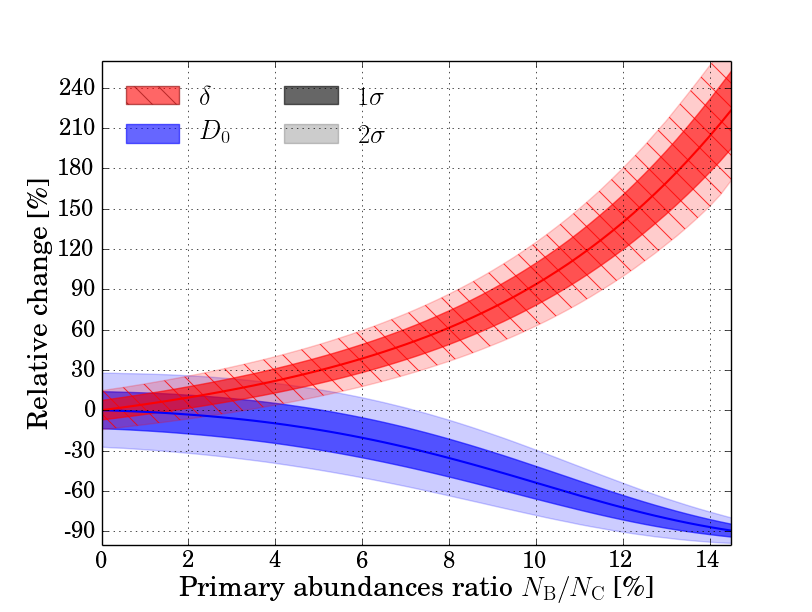}
\includegraphics[width=0.5\textwidth]{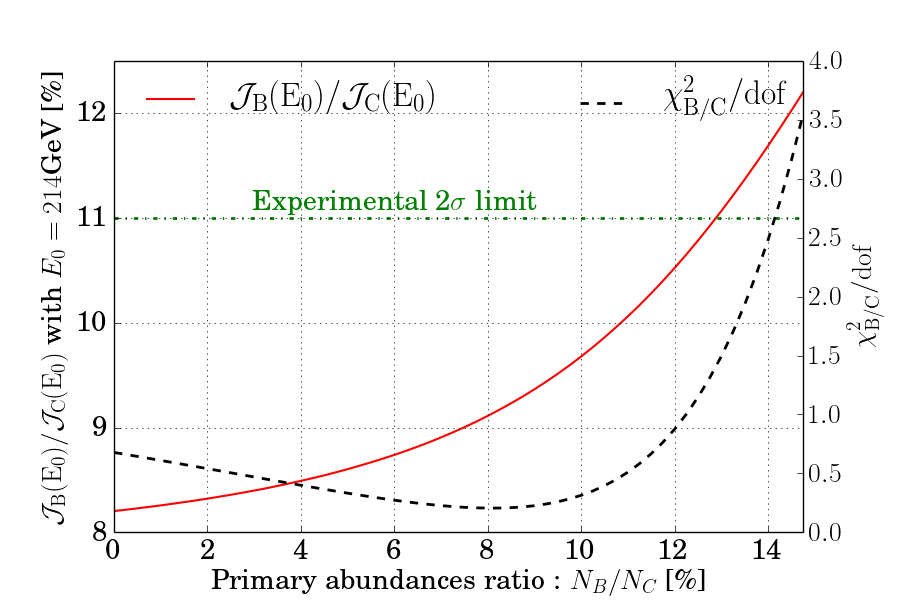}
\caption{
Left panel: variations of the best-fit propagation parameters $D_0$ (blue) and $\delta$ (red) relative to
the benchmark values of Table~\ref{tab:benchmark}, as a function of the primary boron-to-carbon injection ratio.
The reference model corresponds to the conventional no boron hypothesis for which ${N_{\text B}}/{N_{\text C}}$ vanishes.
Right panel: the theoretical value of the B/C ratio at 214\,GeV/nuc  (solid red curve) is plotted as a function of the primary boron-to-carbon injection ratio.
The dashed black curve indicates the goodness of the B/C fit. As long as ${N_{\text B}}/{N_{\text C}}$ does not exceed
{13\%}, the theoretical B/C ratio is within $2\sigma$ from the AMS-02 measurement (dashed-dotted green curve).}
\label{fig:5}
\end{figure}
%FFFFFFFFFFFFFFFFFFFFFFFFFFFFFFFFFFFFFFFFFFFFFFFFFFFFFFFFFFFFFFFFFFFFFFFFFFFFFFFFFFFFF

Apparently little attention has been paid to the bias introduced by the ansatz $Q_{\text B}=0$.
To the best of our knowledge, we quantify it here for the first time. As can be inferred from Eq.~(\ref{eq:b_to_c_primary_B}),
in the presence of a primary source $Q_\text{B}$, the B/C ratio exhibits a plateau as soon as the cross-section ratio
${\sigma_{{\text C} \to {\text B}}}/({\sigma^{\rm diff} + \sigma_{\text B}})$ becomes negligible with respect to the
primary abundances ratio ${N_{\text B}}/{N_{\text C}}$. This happens at sufficiently high energy since $\sigma^{\rm diff}$
increases with the diffusion coefficient $D$. The height of this high-energy B/C plateau is approximately given by the value of
${N_{\text B}}/{N_{\text C}}$.
In the presence of this behaviour, the spectral index $\delta$ must increase to keep fitting the data at low energy,
that is, here around 10\,GeV/nuc. This also implies that $D_0$ decreases with ${N_{\text B}}/{N_{\text C}}$ as
a result of the above-mentioned anti-correlation between the diffusion parameters.

\vskip 0.1cm
We have thus scanned the boron-to-carbon ratio at the source and studied the variations of the best-fit values of $D_0$ and $\delta$
with respect to the reference model of Table~\ref{tab:benchmark}. Our results are illustrated in Fig.~\ref{fig:5}, where the left panel features the
confidence levels for $\delta$ (red) and $D_0$ (blue) as a function of the ${N_{\text B}}/{N_{\text C}}$ ratio.
The B/C fit is particularly sensitive to the last few AMS-02 points, notably the penultimate data point, around 214 GeV/nuc, for which
the B/C ratio is found to be $\sim 9$\%.
In the right panel, the theoretical expectation for that point is plotted (solid red curve) as a function of the primary abundances ratio,
while the dashed black curve indicates how the goodness of fit varies. It is interesting to note that a minor preference is shown for a non-vanishing
fraction of primary boron, around 8\%, due to the marginal preference for a flattening of the ratio already mentioned in the previous section. The ${N_{\text B}}/{N_{\text C}}$ ratio is only loosely constrained to be below 13\%. Such a loose constraint would nominally mean that a spectral index $\delta$ more than
three times larger than its benchmark value would be allowed, with a coefficient $D_0$ one order of magnitude smaller than indicated
in Table~\ref{tab:benchmark}.
In fact, such changes are so extreme that they would clash with other phenomenological or theoretical constraints and should probably be
considered as unphysical. A spectral index $\delta$ in excess of 0.9, corresponding to a relative increase of 100\% with respect to our
benchmark model, is already so difficult to reconcile with the power-law spectrum of nuclei and the present acceleration schemes that
it would probably be excluded. The message is quite remarkable however. The degeneracy of the diffusion parameters with a possible admixture
of primary boron is so strong that it dramatically degrades our capability of determining the best-fit values of $D_0$ and $\delta$, and beyond
them the properties of turbulence, unless other priors are imposed.

%%%%%%%
\section{Cross-section modelling}\label{sigmas}
\begin{figure}[!ht]
\centering
\includegraphics[width=0.5\textwidth]{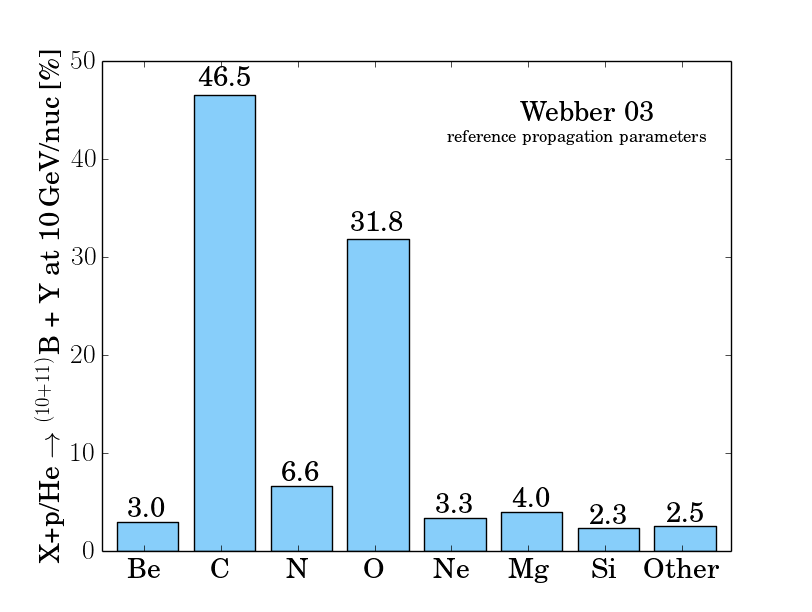}
\caption{Contribution of the various primary nuclear species to the secondary boron flux at $10$~GeV/nuc, as estimated with the semi-empirical code by Webber 03.}
\label{fig:6}
\end{figure}

%%%%%%%
The outcome of cosmic-ray propagation strongly depends on the values of the nuclear production $\sigma_{b \to a}$ and destruction $\sigma_{a}$ cross-sections with the ISM species, mainly protons and helium nuclei. Some of these are measured, albeit in a limited dynamical range, while a significant number of them rely on relatively old semi-empirical formulas, calibrated to the few available data points. In this section, we discuss how parametric changes in these inputs reflect on the B/C ratio. The effect of cross-section systematics was already studied by\citetads{2010A&A...516A..67M}, who parameterised it in terms of a systematic shift with respect to the energy. Since we consider here only the high-energy limit, we simply allowed for a rescaling of the cross-sections. However, we distinguished between two cases: a correlated ($\nearrow \nearrow$) or anti-correlated ($\nearrow \searrow$) rescaling between the production $\sigma_{b \to a}$ and the destruction $\sigma_{a}$ cross-sections. These in fact are not affected by the same uncertainties. It is often the case that the latter are known to a better precision then the former since they rely on a richer set of data. A priori, it is conceivable that several relevant production cross-sections might be varied independently. It is worth noting, however, that only a few nuclei -- notably oxygen and carbon ($\sim 80\%$), and to a lesser extent nitrogen ($\sim 7\%$) -- are in fact responsible for most of the produced boron, as shown in Fig.~\ref{fig:6}.

First, we need to assess the reasonable range over which the various cross-sections of the problem are expected to vary. For this,
we compared our reference models for the destruction and production cross-sections with those used in popular numerical propagation codes such as GALPROP\citepads{2001AdSpR..27..717S} and DRAGON\citepads{2008AIPC.1085..380E}\footnote{Updated version of these two codes can be found at:\url{https://sourceforge.net/projects/galprop} and \url{http://www.dragonproject.org/Home.html},respectively.}. The database implemented these two codes traces back to the GALPROP team work and is based on a number of references including -- but not limited to -- Nuclear Data Sheets and Los Alamos database\citepads{1998nucl.th..12071M} (see\citetads{2001ICRC....5.1836M} and\citetads{2003ICRC....4.1969M} for a more complete list of references). In this work we compare the values given directly by the default cross-section parameterisations without any renormalisation (which can be implemented however).

\begin{figure}[!ht]
\centering
\includegraphics[width=0.5\textwidth]{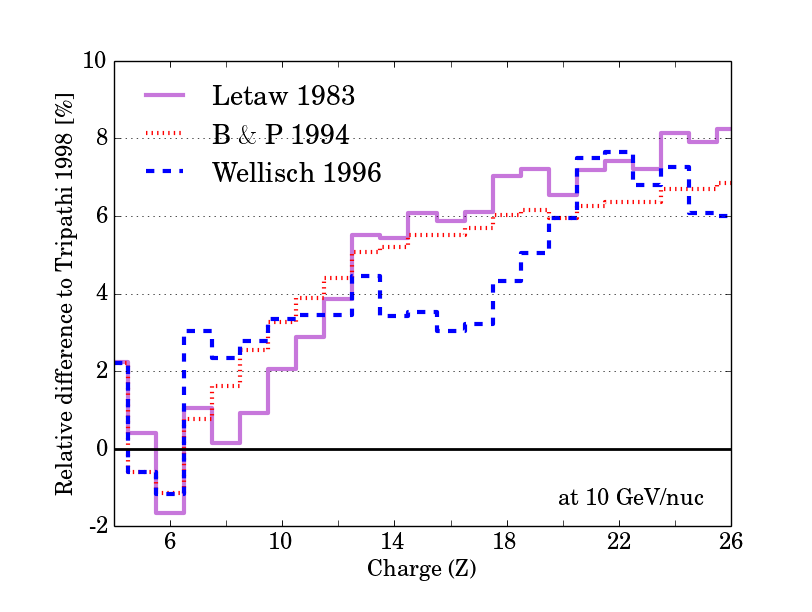}
 \caption{Relative differences between our reference model\citepads{1997lrc..reptQ....T} for the destruction cross-sections $\sigma_{a}$
and the other parameterisations by Letaw 1983\citepads{1983ApJS...51..271L}, Wellish 1996\citepads{1996PhRvC..54.1329W} and
B\&P 1994 \citep{Barashenkov:5725} are displayed as a function of the nucleus charge, at an energy of 10\,GeV/nuc. Each bin is
characterised by a given charge $Z$ and encodes the arithmetic mean over the corresponding isotopes. Only the elements involved
in the cascade from iron to beryllium are displayed.}\label{fig:7}
\end{figure}

\vskip 0.1cm
In the case of the destruction cross-sections $\sigma_{a}$, we compared our reference model \citepads{1997lrc..reptQ....T} with the parameterisations of \citet{Barashenkov:5725}, \citetads{1983ApJS...51..271L} and \citetads{1996PhRvC..54.1329W}. The last case only applies to elements with $Z>5$, while the \citetads{1983ApJS...51..271L} modelling is conserved for lighter nuclei.
Figure~\ref{fig:7} shows the relative differences between our reference model and the three other semi-empirical approaches and allows deriving an indicative lower limit on the systematic uncertainties for the destruction cross-sections of roughly 2 to 10\% for the B/C ratio. The systematic difference is at the 3\% level for the channels (CNO) that contribute most to secondary boron production. The difference to our reference model is stronger for larger charges ($Z > 10$), but these nuclei have a negligible contribution to the B/C ratio.

\vskip 0.1cm
\begin{figure}[!htp]
\includegraphics[width=0.5\textwidth]{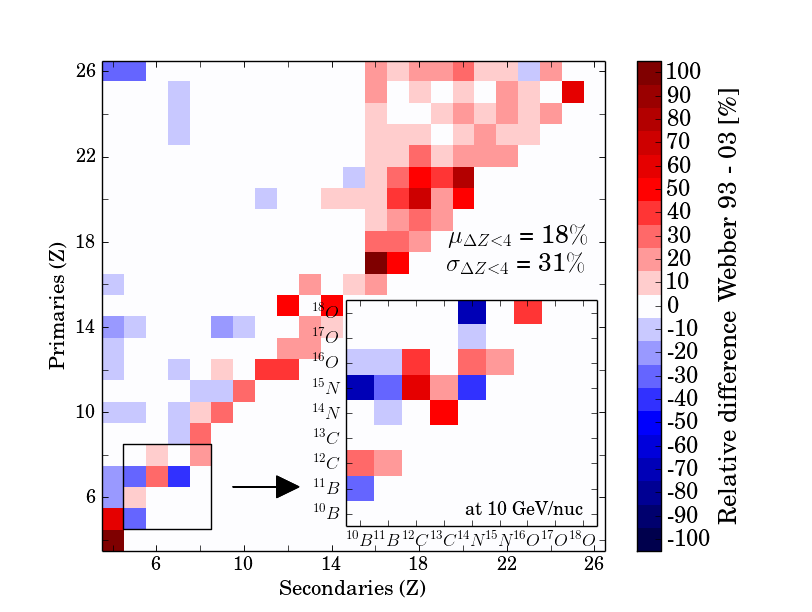}
\includegraphics[width=0.5\textwidth]{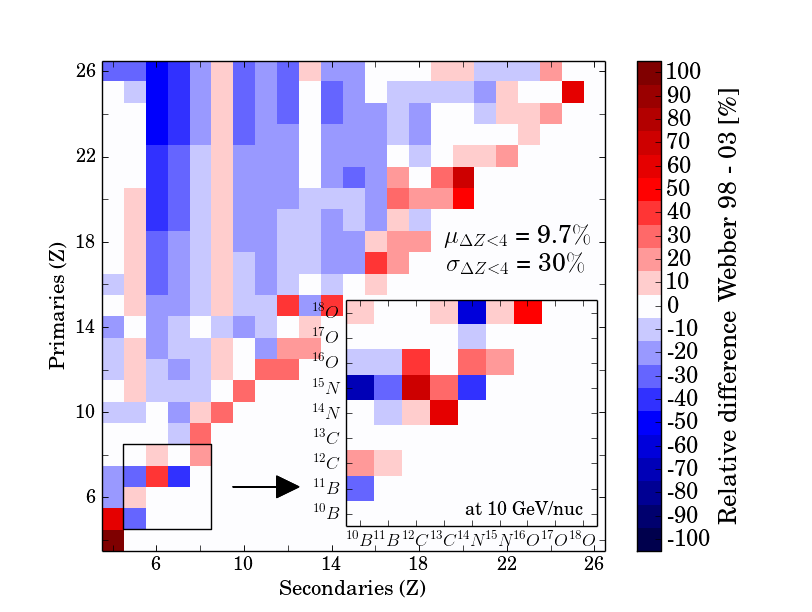}
\includegraphics[width=0.5\textwidth]{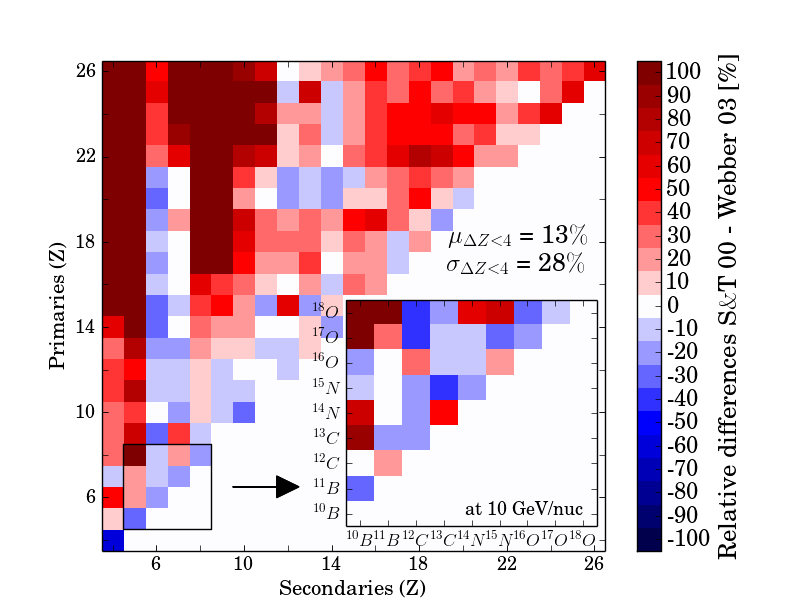}
 \caption{
2D histograms feature the relative differences between various semi-empirical models currently used to calculate the production cross-sections $\sigma_{b \to a}$. Our reference model is \citetads{2003ApJS..144..153W} (Webber 03), and we compare it to the parameterisations from\citetads{1990PhRvC..41..566W} (Webber 93), \citetads{1998PhRvC..58.3539W} (Webber 98) and\citetads{1998ApJ...501..911S} (S\&T 00).
The charges of the parent and child nuclei are given on the vertical and horizontal axes, respectively. The relative difference in each bin is given by the arithmetic mean over the various isotopes of each element. A detailed view provides the most important channels for the B/C ratio studies. For a fragmentation of $\Delta Z<4$, we also give the first and second moments of the uncertainty distributions.}\label{fig:8}
\end{figure}

For the production cross-sections $\sigma_{b \to a}$, one may chose between the semi-empirical approach proposed by\citetads{1998ApJ...501..911S}, subsequently revised in 2000 and called here S\&T 00, and the parameterisation provided by\citetads{1990PhRvC..41..566W} (hereafter Webber 93) and its updates of 1998\citepads{1998PhRvC..58.3539W} and 2003\citepads{2003ApJS..144..153W}. We selected the last set of values as our reference model, to which
we have compared the other parameterisations to gauge the uncertainties that affect, on average, the values of $\sigma_{b \to a}$.
The relative differences between Webber 93, Webber 98, and S\&T 00 with respect to Webber 03 are plotted in the form of the three histograms
of Fig.~\ref{fig:8}. The charges of the parent and child nuclei are given on the vertical and horizontal axes.
The most important reactions, whose cross-sections are higher, correspond to a change of charge $\Delta Z$ not in excess of 3 during the fragmentation process and are located close to the diagonals of the 2D-grids of Fig.~\ref{fig:8}. We first note that the Webber 93 and 98 production cross-sections are on average larger than the values of the Webber 03 reference model. Most of the pixels on the diagonals of the corresponding histograms are red, and we measured an excess $\mu$ on the reactions for which $\Delta Z < 4$ of 18\% and 9.7\% for Webber 93 and Webber 98 as compared to Webber 03. Furthermore, in both cases the dispersion of these differences is quite large and amounts to 31\% for Webber 93 and 30\% for Webber 98. A rapid comparison between S\&T~00 and Webber 03 would also leave the impression that in the former case, the reactions in the upper left corner of the histogram have cross-sections exceedingly larger than for the Webber 03 parameterisation. A close inspection along the diagonal indicates, on the contrary, that the S\&T 00 values for $\Delta Z < 4$ are on average 13\% higher than for the reference model, with a dispersion $\sigma$ of 28\% similar to the other cases.
The main production channels of secondary boron are listed in Table~\ref{tab:sigma_prod_B} and are also displayed in the expanded views of the small square regions that sit in the lower left corners of the histograms of Fig.~\ref{fig:8}. The most relevant reactions involve the stable isotopes of carbon, nitrogen, and oxygen fragmenting into $^{10}$B and $^{11}$B, and are indicated in boldface in Table~\ref{tab:sigma_prod_B}. The largest contributor to secondary boron is $^{12}$C.
The three semi-empirical models with which we compared our Webber 03 reference parameterisation tend to predict production cross-sections that are 15\% for S\&T 00 to 25\% for Webber 93 larger. In contrast, those models underpredict the spallation of $^{16}$O by 10\% in the case of Webber 93 and 98 to 18\% for S\&T 00. In the latter case, the production cross-section of $^{10}$B from $^{14}$N is 68\% larger than for Webber 03. But nitrogen only contributes $\sim$ 7\% of the secondary boron, and this has no significant impact.
To summarise this discussion, the production cross-sections $\sigma_{b \to a}$ can be varied up or down by a factor of order 10-20\% with respect to Webber 03.

\begin{table*}
\centering
\begin{tabular}{|c|c|c|c|c|}
  \hline
   Main production channels   & Webber 03 & S\&T 00 & Webber 98 & Webber 93 \\
   & reference model (RM) & rel. difference to RM & rel. difference to RM & rel. difference to RM \\
   $\sigma_{\text{CNO $\to$ B}}$ at $10~\rm GeV/nuc$ & [mb] &  [\%] &  [\%] &  [\%] \\
  \hline \hline
  \boldmath $\sigma \left(^{12}_{6}\bf{C}\to\ ^{10}_{5}\bf{B} \right)$ & 14.0 & -2.14 & 21.8 & 25.4\\
  \boldmath $\sigma \left(^{12}_{6}\bf{C}\to\ ^{11}_{5}\bf{B} \right)$ & 47.0 & 15.3 & 14.8 & 18.4\\
  $\sigma \left(^{13}_{6}\text{C}\to\ ^{10}_{5}\text{B} \right)$ & 4.70& 92.0 & -2.06 & -0.03\\
  $\sigma \left(^{13}_{6}\text{C}\to\ ^{11}_{5}\text{B} \right)$ & 40.0 & -20.6 & 2.21 & 4.20\\
  \hline
  \boldmath $\sigma \left(^{14}_{7}\bf{N}\to\ ^{10}_{5}\bf{B} \right)$ & 9.90 & 68.1 & 0.14 & 1.01\\
  \boldmath $\sigma \left(^{14}_{7}\bf{N}\to\ ^{11}_{5}\bf{B} \right)$ & 27.2 & 1.33 & -8.86 & -11.0\\
  $\sigma \left(^{15}_{7}\text{N}\to\ ^{10}_{5}\text{B} \right)$ & 9.20 & -6.55 & -70.9 & -70.7\\
  $\sigma \left(^{15}_{7}\text{N}\to\ ^{11}_{5}\text{B} \right)$ & 28.0 & -0.33 & -27.4 &-27.9 \\
  \hline
  \boldmath $\sigma \left(^{16}_{8}\bf{O}\to\ ^{10}_{5}\bf{B} \right)$ & 10.7 & -18.0 & -7.67 & -8.85\\
  \boldmath $\sigma \left(^{16}_{8}\bf{O}\to\ ^{11}_{5}\bf{B} \right)$ & 24.0 & 2.94 & -9.36 &-10.9\\
  $\sigma \left(^{17}_{8}\text{O}\to\ ^{10}_{5}\text{B} \right)$ & 3.60 & 124 & 0.27 & -1.00\\
  $\sigma \left(^{17}_{8}\text{O}\to\ ^{11}_{5}\text{B} \right)$ & 19.7 & 27.3 & 1.42 & -0.09 \\
  $\sigma \left(^{18}_{8}\text{O}\to\ ^{10}_{5}\text{B} \right)$ & 0.70 & 545 & 4.43 & 4.48 \\
  $\sigma \left(^{18}_{8}\text{O}\to\ ^{11}_{5}\text{B} \right)$ & 12.0 & 113 & 2.20 & 0.77 \\
  \hline
\end{tabular}
\caption{
Comparison between different cross-section parameterisations for the main production channels of secondary boron.
The reference model used in our calculations of the fluxes is adapted from\citetads{2003ApJS..144..153W}
(Webber 03) and is compared to previous releases by\citetads{1990PhRvC..41..566W} (Webber 93)
and\citetads{1998PhRvC..58.3539W} (Webber 98) as well as to the work from\citetads{1998ApJ...501..911S}
(S\&T 00). The dominant production channels, which involve the stable isotopes of carbon, nitrogen, and oxygen,
are listed in boldface.}
\label{tab:sigma_prod_B}
\end{table*}

\begin{figure}[!htp]
\includegraphics[width=0.5\textwidth]{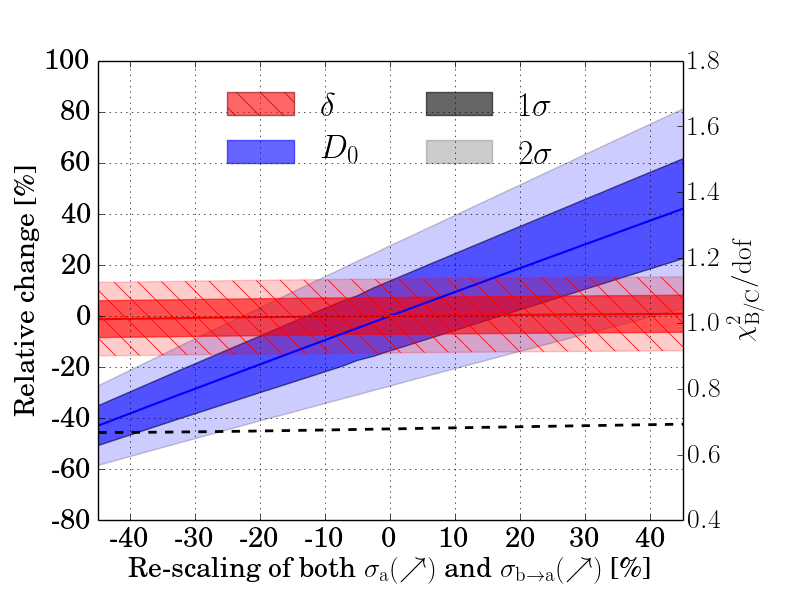}
\includegraphics[width=0.5\textwidth]{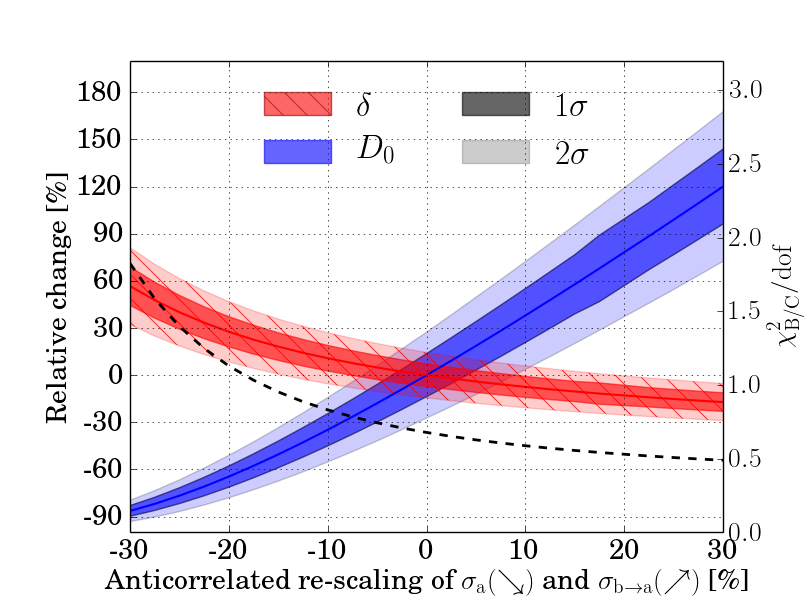}
\caption{Effect of rescaling nuclear cross-sections for boron production channels and destruction ones: the left panel assumes correlated, the right panel anti-correlated rescaling. }\label{fig:9}
\end{figure}

\vskip 0.1cm
Varying the various production and destruction cross-sections has an effect on the calculation of the B/C ratio and thus affects the determination of the propagation parameters $D_{0}$ and $\delta$. Before gauging this effect, we remark that secondary boron is essentially produced by CNO nuclei, as indicated in Fig.~\ref{fig:6}. These are essentially primary species for which ${{\cal J}_b}$ is approximately given by the ratio ${Q_b}/(\sigma^{\rm diff} + \sigma_{b})$ and is proportional to the injection normalisation $N_b$. Furthermore, the relevant destruction cross-sections $\sigma_\text{C}$, $\sigma_\text{N}$ and $\sigma_\text{O}$ being approximately equal to each other, with an effective value ranging from 290 to 317\,mb, we conclude that the flux ratios ${{\cal J}_{b}}/{{\cal J}_\text{C}}$ are given by the corresponding ratios ${N_b}/{N_\text{C}}$ of the injection normalisation constants, with the consequence that relation~(\ref{eq:b_to_c}) simplifies to
\begin{equation}
\frac{{\cal J}_\text{B} (E_k)}{{\cal J}_\text{C} (E_k)} \simeq
\sum_{ Z_b \ge Z_\text{C}}^{ Z_\text{max}}
\frac{\sigma_{b \to \text{B}}}{\sigma^{\rm diff} + \sigma_\text{B}} \cdot
\frac{N_{b}}{N_\text{C}} .
\label{eq:b_to_c_s1}
\end{equation}
As mentioned at the beginning of this section, we first rescaled in our code all production $\sigma_{b \to a}$ and destruction $\sigma_{a}$ cross-sections
by the same amount $\kappa$, which ranges from 0 to 2, to study how $D_{0}$ and $\delta$ are affected by this change. The results
are summarised in the left panel of Fig.~\ref{fig:9}. The diffusion index $\delta$ does not suffer any change, whereas the diffusion normalisation $D_{0}$
increases linearly with the rescaling factor $\kappa$. Multiplying both $\sigma_{b \to \text{B}}$ and $\sigma_\text{B}$ by the same factor $\kappa$ in
Eq.~(\ref{eq:b_to_c_s1}) amounts to dividing the diffusion cross section $\sigma^{\rm diff}$ by $\kappa$. The B/C ratio depends then on the ratio
${\sigma^{\rm diff}}/{\kappa}$, which scales as ${D_{0}}/{\kappa}$. The theoretical prediction on the B/C ratio is not altered as long as that ratio is
kept constant, hence the exact scaling of $D_{0}$ with $\kappa$ displayed in the left panel of Fig.~\ref{fig:9}. The energy behaviour of the B/C ratio
is not sensitive to the rescaling factor $\kappa$, which has been absorbed by $D_{0}$, and the fit yields the same spectral index $\delta$ irrespective
of how much the cross-sections have been changed. Despite the relatively modest alterations, the effect discussed here has two qualitatively interesting
consequences. To commence, a systematic uncertainty on the central value of $D_0$ at the 5 to 10\% level seems unavoidable due to the current uncertainty
level of about 10\% on the nuclear cross-sections. Then, fully correlated changes in both production and destruction cross-sections can break
the degeneracy between $D_0$ and $\delta$.

\vskip 0.1cm
We now analyse the effects of an anti-correlated change of the production $\sigma_{b \to a}$ and destruction $\sigma_{a}$ cross-sections.
Surprisingly, this has never been considered before, as far as we know, although the potential effect of this rescaling clearly is very strong.
Multiplying $\sigma_{b \to \text{B}}$ by a factor $\kappa$ while rescaling $\sigma_\text{B}$ by a complementary factor of $(2 - \kappa)$ leads to the
B/C ratio
\begin{equation}
\frac{{\cal J}_\text{B} (E_k)}{{\cal J}_\text{C} (E_k)} =
\sum_{ Z_b \ge Z_\text{C}}^{ Z_\text{max}}
\left\{
\frac{\sigma_{b \to \text{B}}}{(\sigma^{\text{diff}} + 2 \sigma_\text{B})/\kappa - \sigma_\text{B}}
\right\}
\frac{N_{b}}{N_\text{C}}.
\label{eq:b_to_c_s2}
\end{equation}
Keeping the B/C ratio constant while increasing $\kappa$ at a given energy translates into keeping the ratio
\begin{equation}
\frac{\sigma^{\text{diff}} + 2 \sigma_\text{B}}{\kappa} =
\frac{C E^\delta + 2 \sigma_\text{B}}{\kappa}
\end{equation}
roughly constant, where $C$ is a constant directly proportional to $D_0$. It can be immediately inferred that, when $\kappa$ increases, $C$ and $D_{0}$ have to increase and
thus $\delta$ has to decrease. This trend is confirmed in the right panel of Fig.~\ref{fig:9}. From realistic assessments of the minimum systematic
uncertainties of about 10\% derived from the different cross-section models, we estimate a systematic uncertainty of 10\% on $\delta$ and of 40\% on
$D_{0}$.

%%%%%%%%
\section{Systematics related to CR propagation modelling}\label{propmodeling}
%%%%%%%%
A significant effort has been made in recent years to provide increasingly sophisticated modelling of the CR diffusion environment, source distribution, and alternative forms of CR transport. In this section we discuss a perhaps surprising
conclusion: these effects are less relevant for the prediction of B/C than the effects discussed previously (which are instead usually neglected)!
The message is: although the efforts invested by the community in refining CR propagation modelling could have and
have had important implications for other observables, for the mere purpose of fitting B/C to infer diffusion propagation parameters  they are to a large extent unnecessary complications, until one can significantly reduce the biases previously discussed.
%%%%%%%%
\subsection{Geometric effects}
\begin{figure}[!ht]
\centering
\includegraphics[width=0.45\textwidth]{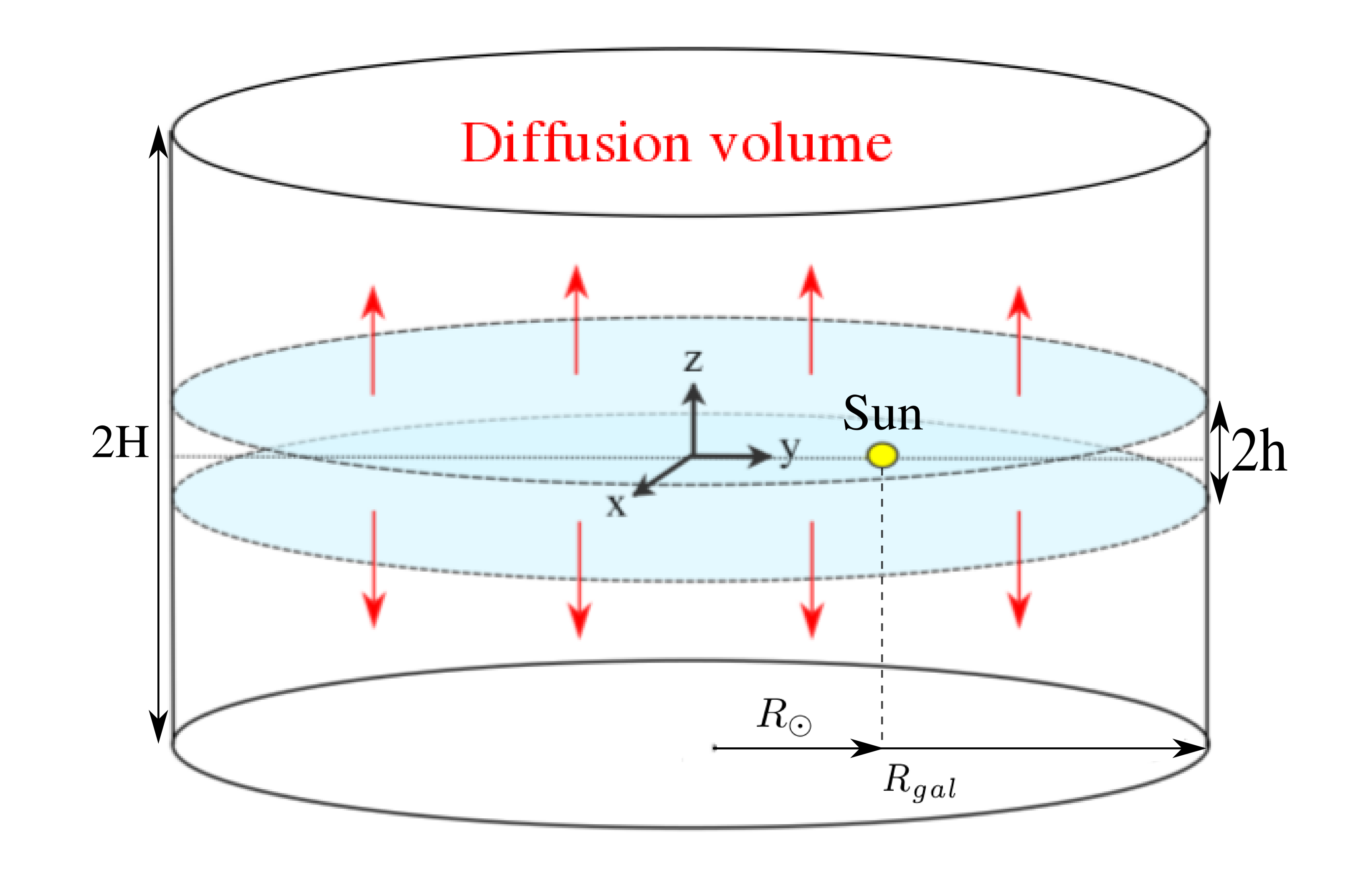}
\caption{Cylindric model: the matter is homogeneously distributed inside a thin disk of thickness $2h$ and radius $R_{gal}=20$\,kpc. The solar system is at $R_\odot\simeq 8$\,kpc from the Galactic centre.}\label{fig:10}
\end{figure}

%%%%%%%%
The crude modelling of the diffusive halo as an infinite slab may appear too simplistic. In this section, we estimate the
effects of a 2D cylindrical diffusion box, modelled as in Fig.~(\ref{fig:10}). Furthermore, we assess the effect
of adding a radial dependence in the injection term, as opposed to the uniform hypothesis. These can be seen as upper
limits to reasonable systematics due to simplified description of the spatial dependence of the diffusion medium or source term:
given our limited knowledge on this subject, even the most detailed modelling of the propagation medium and source term, in fact, may not be fully realistic.

The formalism in such a situation is well known and we do not repeat it here (it has been summarised for instance in ~\cite{Putze:2010zn}). It suffices to say that to take advantage of the cylindrical symmetry, Eq.(\ref{eqprop2}) can be projected on the basis of the zero order Bessel functions $J^i_0(r)=J_0\left(\xi_i\frac{r}{R_{gal}}\right)$ ensuring that the density vanishes on the edge of the cylinder of radius $R_{gal}=20$\,kpc. The flux of an isotope is then the sum over all its harmonic components
\begin{equation}
{\cal J}_{a}(E_k,R_\odot)=\sum_i^\infty J_0\left(\xi_i\frac{R_\odot}{R_{Gal}}\right){\cal J}_{a}^{i}(E_k)\,.
\end{equation}

The results, reported in Table~\ref{tab:geometry}, allow us to draw a few conclusions:
\begin{itemize}
\item the presence of a new escape surface at $R_{\rm gal}\simeq 20\,$kpc is basically irrelevant:
the best-fit $\delta$ and its error remain the same, with a statistically insignificant, 2\% modification of
the best-fit value of $D_0$;
\item perhaps more surprisingly, even the replacement of a uniform source distribution with a commonly assumed donut distribution of the form\citepads{2004A&A...422..545Y}
\begin{equation}
q(r)\propto \displaystyle\left(\frac{r+0.55}{R_\odot+0.55}\right)^{1.64} \exp\left(-4.01\left(\frac{r-R_\odot}{R_\odot+0.55}\right)\right)
\end{equation}
has minor effects, a mere 1\% modification in the best-fit determination of $\delta$,
and a $\sim 13\%$ lowering of the best-fit value of $D_0$, still statistically insignificant (roughly a 1$\,\sigma$ effect);
\item since the goodness of fit is similar, the B/C observable is essentially insensitive to these improvements.
Unless they are justified by the goal of matching or predicting other observables, the complication brought by the 2D
modelling of the problem are unnecessary in achieving a good description of the data.
\end{itemize}

\begin{table*}
	\centering
\begin{tabular}{|l|c|c|c|}
  \hline
   Geometry & Plane -- 1D & Cylindrical -- 2D  & Cylindrical -- 2D  \\
   & & \small homogeneous source distribution & \small realistic source distribution \\
  \hline \hline

  $D_{0}$ [kpc$^2$/Myr] & $(5.8\pm0.7)\cdot10^{-2}$ & $(5.7\pm0.7)\cdot10^{-2}$ & $(5.0\pm0.6)\cdot10^{-2}$\\

  $\Delta D_{0}^{\text{1D}}/D_{0}^{\text{1D}}$ & N/A    & $-2\%$            & $-13\%$\\ [2mm]

  $\delta$                    & $0.441\pm0.031$        & $0.439\pm0.031$        & $0.445\pm0.032$ \\

  $\Delta \delta^{\text{1D}}/\delta^{\text{1D}}$ & N/A    & $0\%$            & $+1\%$\\[2mm]

  $\chi^2_{\text{B/C}}$/ndof & $5.4/8\approx0.68$ & $5.4/8\approx0.68$ & $5.5/8\approx0.69$\\
  \hline
\end{tabular}
\caption{Results on the propagation parameters fitted on the B/C for different geometries.}\label{tab:geometry}
\end{table*}

%%%%%%
\subsection{Convective wind}
%%%%%%
Although the high-energy CR propagation is mostly diffusive, the advection outside the Galactic plane (for instance due to
stellar winds) has a non-negligible effect, which we now quantify. We adopted the simplest model of constant velocity wind, directed outside the galactic plane, with magnitude $u$. Taking this effect into account this effect, the 1D, stationary propagation equation can be written as
\begin{align}
\nonumber &-\frac{\partial}{\partial z} \left( D\frac{\partial}{\partial z} \psi_a \right) + \frac{\partial}{\partial z} \left(u \psi_a \right) -\frac{\partial}{\partial E} \left( \frac{1}{3}\frac{d u}{dz} E_k\frac{(E_k+2m)}{E_k+m} \psi_a \right) \\ &+ \delta(z)\sigma_{a} v \frac{\mu}{m_\text{ISM}} \psi_a
= 2h \delta(z) q_{a} + \delta(z) \sum_{Z_b \geqslant Z_a}^{Z_\text{max}} \sigma_{b\to a} v \frac{\mu}{m_\text{ISM}} \psi_b ,
\label{eq:propwind}
\end{align}
The two new terms (second and third one on the LHS) account for the advection of the cosmic-ray density and the adiabatic losses, respectively. A characteristic time of these two processes can be estimated inside the thin disk of matter :
\begin{align}
\tau_\text{advection}&=\frac{h}{u}=\frac{0.1\,\text{kpc}}{20\,\text{km/s}} \nonumber \\&=5 \cdot \left(\frac{h}{0.1\,\text{kpc}}\right) \cdot \left(\frac{20\,\text{km/ s}}{u}\right)\,\text{My},
\end{align}
and
\begin{align}
 \tau_\text{adiabatic} &= \left(\frac{1}{3A}\left(\nabla  u\right)\right)^{-1}\simeq 3A\frac{h}{u} \nonumber \\ &\approx 15\cdot \left(\frac{h}{0.1\,\text{kpc}}\right)\cdot \left(\frac{20\,\text{km/s}}{u}\right) \cdot \left(\frac{A}{1}\right)\,\text{My}.
 \end{align}
 This means that adiabatic losses can be safely neglected compared to the typical diffusion time of
 \begin{align}
 \tau&_\text{diffusion}\left({\cal R} > 10\,\text{GV}\right)< \tau_\text{diffusion} (10\,\text{GV}) = \frac{h\,H}{D (10\,\text{GV})} \nonumber \\ &= 2 \cdot \left(\frac{h}{0.1\,\text{kpc}}\right) \cdot \left(\frac{H}{4\,\text{kpc}}\right) \cdot \left(\frac{5.8 \cdot 10^{-2}\cdot (2 \cdot 10)^{0.44}\,\text{kpc$^2$/My}}{D}\right)\,\text{My}.
 \end{align}
It is clear that our previous results  provide a suitable first-order approximation at least at high energy, with the leading correction at energies
near 10\,GeV/n  given especially by the advection. The adiabatic energy loss, instead, is several times smaller and can be safely ignored in the following.

\begin{figure}[!ht]
\centering
\includegraphics[width=0.5\textwidth]{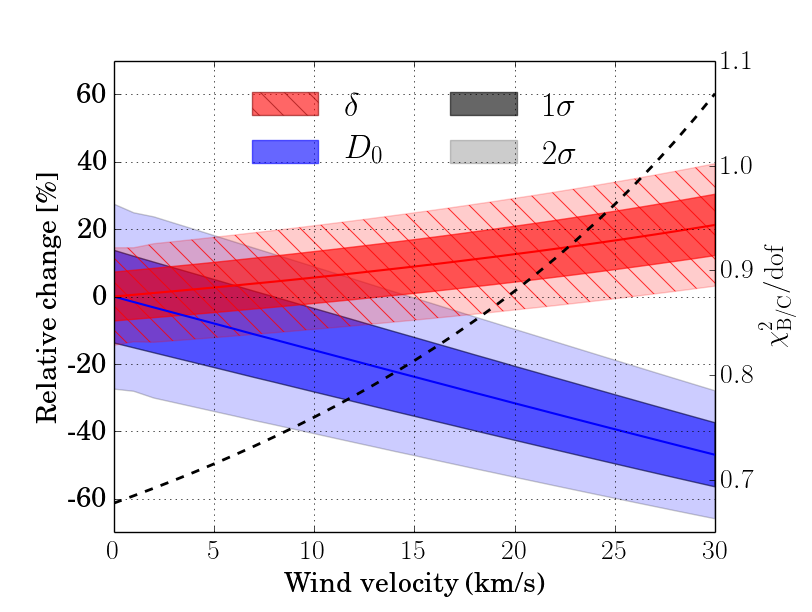}
\caption{Variations of the best-fit propagation parameters with respect to the velocity of the convective wind.}\label{fig:11}
\end{figure}

The solution of Eq.~(\ref{eq:propwind}) neglecting adiabatic losses has the same form of  Eq.~(\ref{eq:flux}) for the flux of stable species, modulo the change
\begin{equation}
D\rightarrow D'=\frac{H\,u}{1-\exp\left(-\frac{H\,u}{D}\right)},\label{dprime}
\end{equation}
so that the behaviour of the solution smoothly interpolates
between the convective timescale at low energy and the diffusive one at high energy: this can be simply checked by neglecting the exponential with respect to unity for a high value of its argument, or Taylor-expanding it to first order in the opposite limit. This formula also suggests that, if one fits the data by neglecting the convective wind, one biases its result towards a lower value of $\delta$, and a corresponding higher value of $D$, so to reproduce a flatter dependence with energy at low-energy as for the case described by Eq.~(\ref{dprime}), as illustrated in Fig.~\ref{fig:11}. Quantitatively, a variation of $15\,$km/s in $u$ is roughly similar to a 1$\sigma$ shift in the benchmark parameters. Note, however, that the goodness of the fit worsens, or in other words, high-energy data are better described by a pure diffusive behaviour than by a convective-diffusive one. Overall,
we conclude that these effects appear still somewhat less important in determining the diffusion parameters from high-energy data than the role of primary boron or even cross-section uncertainties. While convection, adiabatic losses, reacceleation, etc. are important to account for when extending the analysis down to very low energies (sub-GeV/nuc) or in global analyses, they do not currently constitute the main limitations to the determination of $D_0$ or $\delta$ from high-energy data.

%%%%%%
\section{Conclusion}\label{conclusions}
%%%%%%

The high-precision measurements of cosmic-ray fluxes that have become available in recent years prompt the question of the theoretical uncertainties inherent to the models used to interpret them. In this article, we have compared
the effect of different theoretical biases with statistical uncertainties in the determination of diffusion parameters
from the boron-to-carbon flux ratio, or B/C. This is representative of a much broader class of observables,
involving ratios of secondary to primary species, which have been recognised as key tools for diagnostics in
cosmic-ray astrophysics. We adopted a pedagogical approach, showing and interpreting the results whenever
possible within simple analytical models. We also used preliminary AMS-02 data and limited the analysis to energies
above 10\,GeV/nuc, which gives a pessimistic---hence conservative---estimate of the statistical uncertainties that will eventually be available.

\begin{table*}
	\centering
\begin{tabular}{|l|c|c|c|c|}
  \hline
   & Wind & 1D/2D geometry& Cross-sections & Primary boron\\
  \hline \hline
  $\Delta D_0/D_{0}$ & $-40\%$ & $-2$ to $-13\%$ & $\pm 60\%$  & $0$ to $-90\%$\\ [2mm]
  $\Delta \delta/\delta$ & $+15\%$  & $0$ to $+1\%$   &  $\pm 20\%$       & $0$ to $+100\%$ \\
  \hline
\end{tabular}
\caption{Summary of the main systematics found in current analyses in determining the propagation parameters by fitting the B/C ratio.}\label{finaltab}
\end{table*}

Our main results, summarised in Table~\ref{finaltab}, are the following:
\begin{itemize}
\item The single most important effect that we quantified (to the best of our knowledge, for the first time) is the degeneracy between diffusion parameters
and a small injection of primary boron at the source, finding at present even a statistically insignificant preference for a small but finite value for a primary boron flux.
This degeneracy cannot be removed by high-precision measurements of B/C, but probably requires multi-messenger tests and certainly demands
further investigations, in particular if data should manifest a significant preference for a high-energy flattening of secondary-to-primary ratios.

\item The second most important theoretical uncertainty is associated to cross-sections. In particular, anti-correlated modifications in the destruction and
production cross-sections with respect to reference values may also have an effect on the determination of the diffusion index $\delta$, another effect
discussed here for the first time. This should be kept in mind when comparing the outcome of data analyses relying on different databases for cross-sections.
The good news is that this problem is not due to intrinsic limitations in the astrophysical modelling or the lack of astrophysical data, but to the scarce laboratory measurements available. For the case of boron, experiments of production cross-sections via spallation of oxygen, carbon and, to a minor extent, nitrogen, are essentially what would be
needed to set the predictions on much firmer grounds.

\item Other effects we tested for are typically less important and are similar to or smaller than statistical uncertainties:
effects such as those of convective winds, certainly important in more complete analyses including low-energy data, appear unlikely to bring uncertainties large enough to compete with the above-mentioned uncertainties. We also showed how the geometry of the diffusive box and the distribution of sources is virtually irrelevant, at least if only a B/C data
analysis is concerned. More or less realistic radial distribution of sources, while it may marginally affect the determination of $D_0$, is still indistinguishable from the goodness-of-fit point of view. Another outcome of this exercise is that at least at the 10\% level, $D_0$ is degenerate with a choice of geometry and source distribution, in addition to the already well-known degeneracy with the diffusive halo height $H$.
\end{itemize}
In conclusion, we found that the main uncertainties in inferring diffusion parameters from B/C (and we expect from other secondary-to-primary ratios, too) depend on theoretical priors on sources (linked to sites and mechanisms of acceleration!) and, to a lesser extent, to nuclear cross-sections. While exploring more complicated schemes and geometries for the diffusion may thus be important, we can anticipate that sensitivity to such effects will probably require fixing more mundane questions first!
A multi-messenger strategy, coupled to a new measurement campaign of nuclear cross-sections, appears to be a next crucial step in that direction.

%%%%%%%%%%%%%%%%%%%%%%%%%%%%%%%%%%%%%%%%%%%%%%%%%%%%%%%%%%
%%%%%%%%%%%%%%%%%%%%%%%%%%%%%%%%%%%%%%%%%%%%%%%%%%%%%%%%%%
%
\vskip 1.0cm
\begin{acknowledgements}
We would like to thank David Maurin for sharing useful data, notably Webber 2003 production cross-sections and expertise on cross-sections.
Part of this work was supported by the French \emph{Institut universitaire de France}, by the French
\emph{Agence Nationale de la Recherche} under contract  12-BS05-0006 DMAstroLHC,
and by the \emph{Investissements d'avenir}, Labex ENIGMASS.
\end{acknowledgements}

%%%%%%%%%%%%%%%%%%%%%%%%%%%%%%%%%%%%%%%%%%%%%%%%%%%%%%%%%%
%%%%%%%%%%%%%%%%%%%%%%%%%%%%%%%%%%%%%%%%%%%%%%%%%%%%%%%%%%
\bibliographystyle{aa}
\bibliography{draft4}

\begin{thebibliography}{30}
\expandafter\ifx\csname natexlab\endcsname\relax\def\natexlab#1{#1}\fi

\bibitem[{33rd Intern. Cosmic Ray~Conf.(2013)}]{2013..ICRC}
33rd Intern. Cosmic Ray~Conf., P., ed. 2013, {Precision Measurement of the
  Cosmic Ray Boron-to-Carbon Ratio with AMS [AMS collaboration]}, ed. P.~33rd
  Intern. Cosmic Ray~Conf.

\bibitem[{{Adriani} {et~al.}(2014){Adriani}, {Barbarino}, {Bazilevskaya},
  {Bellotti}, {Boezio}, {Bogomolov}, {Bongi}, {Bonvicini}, {Bottai}, {Bruno},
  {Cafagna}, {Campana}, {Carbone}, {Carlson}, {Casolino}, {Castellini}, {De
  Pascale}, {De Santis}, {De Simone}, {Di Felice}, {Formato}, {Galper},
  {Giaccari}, {Karelin}, {Kheymits}, {Koldashov}, {Koldobskiy}, {Krut`kov},
  {Kvashnin}, {Leonov}, {Malakhov}, {Marcelli}, {Martucci}, {Mayorov}, {Menn},
  {Mikhailov}, {Mocchiutti}, {Monaco}, {Mori}, {Munini}, {Nikonov}, {Osteria},
  {Papini}, {Pearce}, {Picozza}, {Pizzolotto}, {Ricci}, {Ricciarini},
  {Rossetto}, {Sarkar}, {Simon}, {Sparvoli}, {Spillantini}, {Stozhkov},
  {Vacchi}, {Vannuccini}, {Vasilyev}, {Voronov}, {Wu}, {Yurkin}, {Zampa},
  {Zampa}, \& {Zverev}}]{2014PhR...544..323A}
{Adriani}, O., {Barbarino}, G.~C., {Bazilevskaya}, G.~A., {et~al.} 2014,
  \physrep, 544, 323

\bibitem[{Barashenkov \& Polanski(1994)}]{Barashenkov:5725}
Barashenkov, V.~S. \& Polanski, A. 1994, Electronic guide for nuclear
  cross-sections: version 1994, Tech. Rep. E2-94-417. JINR-E2-94-417, Joint
  Inst. Nucl. Res., Dubna

\bibitem[{{Binns} {et~al.}(2008){Binns}, {Wiedenbeck}, {Arnould}, {Cummings},
  {de Nolfo}, {Goriely}, {Israel}, {Leske}, {Mewaldt}, {Stone}, \& {von
  Rosenvinge}}]{2008NewAR..52..427B}
{Binns}, W.~R., {Wiedenbeck}, M.~E., {Arnould}, M., {et~al.} 2008, \nar, 52,
  427

\bibitem[{Blasi(2009)}]{Blasi:2009hv}
Blasi, P. 2009, Phys.Rev.Lett., 103, 051104

\bibitem[{Blasi \& Serpico(2009)}]{Blasi:2009bd}
Blasi, P. \& Serpico, P.~D. 2009, Phys.Rev.Lett., 103, 081103

\bibitem[{{Evoli} {et~al.}(2008){Evoli}, {Gaggero}, {Grasso}, \&
  {Maccione}}]{2008AIPC.1085..380E}
{Evoli}, C., {Gaggero}, D., {Grasso}, D., \& {Maccione}, L. 2008, in American
  Institute of Physics Conference Series, Vol. 1085, American Institute of
  Physics Conference Series, ed. F.~A. {Aharonian}, W.~{Hofmann}, \&
  F.~{Rieger}, 380--383

\bibitem[{{Letaw} {et~al.}(1983){Letaw}, {Silberberg}, \&
  {Tsao}}]{1983ApJS...51..271L}
{Letaw}, J.~R., {Silberberg}, R., \& {Tsao}, C.~H. 1983, \apjs, 51, 271

\bibitem[{{Lodders}(2003)}]{2003ApJ...591.1220L}
{Lodders}, K. 2003, Apj, 591, 1220

\bibitem[{{Mashnik} {et~al.}(1998){Mashnik}, {Sierk}, {Van Riper}, \&
  {Wilson}}]{1998nucl.th..12071M}
{Mashnik}, S.~G., {Sierk}, A.~J., {Van Riper}, K.~A., \& {Wilson}, W.~B. 1998,
  ArXiv Nuclear Theory e-prints [\eprint{nucl-th/9812071}]

\bibitem[{Maurin {et~al.}(2001)Maurin, Donato, Taillet, \&
  Salati}]{Maurin:2001sj}
Maurin, D., Donato, F., Taillet, R., \& Salati, P. 2001, Astrophys.J., 555, 585

\bibitem[{{Maurin} {et~al.}(2014){Maurin}, {Melot}, \&
  {Taillet}}]{2014A&A...569A..32M}
{Maurin}, D., {Melot}, F., \& {Taillet}, R. 2014, \aap, 569, A32

\bibitem[{{Maurin} {et~al.}(2010){Maurin}, {Putze}, \&
  {Derome}}]{2010A&A...516A..67M}
{Maurin}, D., {Putze}, A., \& {Derome}, L. 2010, \aap, 516, A67

\bibitem[{Maurin {et~al.}(2002)Maurin, Taillet, Donato, Salati, Barrau,
  {et~al.}}]{Maurin:2002ua}
Maurin, D., Taillet, R., Donato, F., {et~al.} 2002
  [\eprint[arXiv]{astro-ph/0212111}]

\bibitem[{Mertsch \& Sarkar(2009)}]{Mertsch:2009ph}
Mertsch, P. \& Sarkar, S. 2009, Phys.Rev.Lett., 103, 081104

\bibitem[{{Mertsch} \& {Sarkar}(2014)}]{2014PhRvD..90f1301M}
{Mertsch}, P. \& {Sarkar}, S. 2014, Prd, 90, 061301

\bibitem[{{Moskalenko} \& {Mashnik}(2003)}]{2003ICRC....4.1969M}
{Moskalenko}, I.~V. \& {Mashnik}, S.~G. 2003, International Cosmic Ray
  Conference, 4, 1969

\bibitem[{{Moskalenko} {et~al.}(2001){Moskalenko}, {Mashnik}, \&
  {Strong}}]{2001ICRC....5.1836M}
{Moskalenko}, I.~V., {Mashnik}, S.~G., \& {Strong}, A.~W. 2001, International
  Cosmic Ray Conference, 5, 1836

\bibitem[{{Ptuskin} {et~al.}(1997){Ptuskin}, {Voelk}, {Zirakashvili}, \&
  {Breitschwerdt}}]{1997A&A...321..434P}
{Ptuskin}, V.~S., {Voelk}, H.~J., {Zirakashvili}, V.~N., \& {Breitschwerdt}, D.
  1997, \aap, 321, 434

\bibitem[{Putze {et~al.}(2010)Putze, Derome, \& Maurin}]{Putze:2010zn}
Putze, A., Derome, L., \& Maurin, D. 2010, Astron.Astrophys., 516, A66

\bibitem[{{Silberberg} {et~al.}(1998){Silberberg}, {Tsao}, \&
  {Barghouty}}]{1998ApJ...501..911S}
{Silberberg}, R., {Tsao}, C.~H., \& {Barghouty}, A.~F. 1998, \apj, 501, 911

\bibitem[{{Strong} \& {Moskalenko}(2001)}]{2001AdSpR..27..717S}
{Strong}, A.~W. \& {Moskalenko}, I.~V. 2001, Advances in Space Research, 27,
  717

\bibitem[{Strong {et~al.}(2007)Strong, Moskalenko, \& Ptuskin}]{Strong:2007nh}
Strong, A.~W., Moskalenko, I.~V., \& Ptuskin, V.~S. 2007,
  Ann.Rev.Nucl.Part.Sci., 57, 285

\bibitem[{{Tripathi} {et~al.}(1997){Tripathi}, {Cucinotta}, \&
  {Wilson}}]{1997lrc..reptQ....T}
{Tripathi}, R.~K., {Cucinotta}, F.~A., \& {Wilson}, J.~W. 1997, {Universal
  Parameterization of Absorption Cross Sections}, Tech. rep.

\bibitem[{{Tripathi} {et~al.}(1999){Tripathi}, {Cucinotta}, \&
  {Wilson}}]{1999NIMPB.155..349T}
{Tripathi}, R.~K., {Cucinotta}, F.~A., \& {Wilson}, J.~W. 1999, Nuclear
  Instruments and Methods in Physics Research B, 155, 349

\bibitem[{{Webber} {et~al.}(1990){Webber}, {Kish}, \&
  {Schrier}}]{1990PhRvC..41..566W}
{Webber}, W.~R., {Kish}, J.~C., \& {Schrier}, D.~A. 1990, \prc, 41, 566

\bibitem[{{Webber} {et~al.}(2003){Webber}, {Soutoul}, {Kish}, \&
  {Rockstroh}}]{2003ApJS..144..153W}
{Webber}, W.~R., {Soutoul}, A., {Kish}, J.~C., \& {Rockstroh}, J.~M. 2003,
  Apjs, 144, 153

\bibitem[{{Webber} {et~al.}(1998){Webber}, {Soutoul}, {Kish}, {Rockstroh},
  {Cassagnou}, {Legrain}, \& {Testard}}]{1998PhRvC..58.3539W}
{Webber}, W.~R., {Soutoul}, A., {Kish}, J.~C., {et~al.} 1998, \prc, 58, 3539

\bibitem[{{Wellisch} \& {Axen}(1996)}]{1996PhRvC..54.1329W}
{Wellisch}, H.~P. \& {Axen}, D. 1996, \prc, 54, 1329

\bibitem[{{Yusifov} \& {K{\"u}{\c c}{\"u}k}(2004)}]{2004A&A...422..545Y}
{Yusifov}, I. \& {K{\"u}{\c c}{\"u}k}, I. 2004, \aap, 422, 545

\end{thebibliography}

%%%%%%%%%%%%%%%%%%%%%%%%%%%%%%%%%%%%%%%%%%%%%%%%%%%%%%%%%%
%%%%%%%%%%%%%%%%%%%%%%%%%%%%%%%%%%%%%%%%%%%%%%%%%%%%%%%%%%
\end{document}